%% file: paper.tex
\documentclass[aps,prd,twocolumn,amsfonts,amssymb,amsmath,showpacs,letterpaper,superscriptaddress,nofootinbib,altaffilletter,fontenc]{revtex4-2}

\usepackage[citecolor=blue,linkcolor=blue,colorlinks=true,urlcolor=blue]{hyperref}

\usepackage{color}
\usepackage{graphicx}
\usepackage{latexsym}
\usepackage{float}
\usepackage{amsmath}
\usepackage{amssymb}
\usepackage{multirow}
\usepackage{ifthen}
\usepackage{slashed}
\usepackage{lineno}
\usepackage{makecell} % new line in a table cell
\usepackage[mathscr]{euscript} % for Detection Efficiency E
\usepackage{mathrsfs} % for Detection Efficiency E
\usepackage[normalem]{ulem}
\usepackage{orcidlink}

\newcommand{\hrss}{h_{\textrm{rss}}}

\usepackage[normalem]{ulem}

\newcommand{\makevisible}[1]{\textcolor{red}{#1}}
\newcommand{\switch}[1]{%
  \ifthenelse{\equal{#1}{0}}{\renewcommand{\makevisible}[1]{}}{}}
\switch{1} %-- set to 1 (0) to make visible (invisible)
\def\version$#1,v #2 #3${#2}

\usepackage{array}
\newcolumntype{C}[1]{>{\centering\let\newline\\\arraybackslash\hspace{0pt}}m{#1}}

\input{macros}

\begin{document}

% --- results macros

\title{An Optically Targeted Search for Gravitational Waves emitted by Core-Collapse Supernovae during the Third Observing Run of Advanced LIGO and Advanced Virgo}

%%%%%%%%%%%%%%%%%%%%
% Direct contribution:
\author{Marek J. Szczepa\'nczyk\,\orcidlink{0000-0002-6167-6149}}
\email[E-mail: ]{marek.szczepanczyk@ligo.org}
\affiliation{Department of Physics, University of Florida, Gainesville, FL 32611-8440, USA}
\affiliation{Faculty of Physics, University of Warsaw, Ludwika Pasteura 5, 02-093 Warsaw, Poland}
\author{Yanyan~Zheng}
\affiliation{Institute of Multi-messenger Astrophysics and Cosmology, Missouri University of Science and Technology, Rolla, MO 65409, USA}
\author{Javier~M.~Antelis}
\affiliation{Embry-Riddle Aeronautical University, Prescott, AZ 86301, USA}
\affiliation{Tecnologico de Monterrey, Escuela de Ingeniería y Ciencias, Monterrey, N.L., 64849, México}
\author{Michael~Benjamin}
\affiliation{University of Texas Rio Grande Valley, Brownsville, TX 78539, USA}
\author{Marie-Anne~Bizouard}
\affiliation{Artemis, Universit\'e C\^ote d'Azur, Observatoire de la C\^ote d'Azur, CNRS, F-06304 Nice, France}
\author{Alejandro~Casallas-Lagos}
\affiliation{Embry-Riddle Aeronautical University, Prescott, AZ 86301, USA}
\affiliation{Departamento de F\'isica, Universidad de Guadalajara, Guadalajara, Jal., 44430, M\'exico}
\author{Pablo~Cerd\'a-Dur\'an}
\affiliation{Departamento de Astronom\'{\i}a y Astrof\'{\i}sica, Universitat de Val\`{e}ncia, E-46100 Burjassot, Val\`{e}ncia, Spain}
\affiliation{Observatori Astron\`omic, Universitat de Val\`encia, E-46980 Paterna, Val\`encia, Spain}
\author{Derek~Davis}
\affiliation{LIGO Laboratory, California Institute of Technology, Pasadena, CA 91125, USA}
\author{Dorota~Gondek-Rosi\'nska}
\affiliation{Astronomical Observatory, University of Warsaw, al. Ujazdowskie 4, 00-478 Warsaw, Poland}
\author{Sergey~Klimenko}
\affiliation{Department of Physics, University of Florida, Gainesville, FL 32611-8440, USA}
\author{Claudia~Moreno}
\affiliation{Embry-Riddle Aeronautical University, Prescott, AZ 86301, USA}
\affiliation{Departamento de F\'isica, Universidad de Guadalajara, Guadalajara, Jal., 44430, M\'exico}
\author{Martin~Obergaulinger}
\affiliation{Departamento de Astronom\'{\i}a y Astrof\'{\i}sica, Universitat de Val\`{e}ncia, E-46100 Burjassot, Val\`{e}ncia, Spain}
\author{Jade~Powell}
\affiliation{Centre for Astrophysics and Supercomputing, Swinburne University of Technology, Hawthorn, VIC 3122, Australia}
\affiliation{ARC Centre of Excellence for Gravitational Wave Discovery (OzGrav), Melbourne, Australia}
\author{Dymetris~Ramirez}
\affiliation{Embry-Riddle Aeronautical University, Prescott, AZ 86301, USA}
\author{Brad~Ratto}
\affiliation{Embry-Riddle Aeronautical University, Prescott, AZ 86301, USA}
\author{Colter~Richardson}
\affiliation{University of Tennessee, Knoxville, TN 37996-1200, USA}
\author{Abhinav~Rijal}
\affiliation{Embry-Riddle Aeronautical University, Prescott, AZ 86301, USA}
\affiliation{Institute Of Engineering (Pulchowk Campus), Tribhuvan University, Lalitpur, Nepal}
\author{Amber~L.~Stuver}
\affiliation{Department of Physics, Villanova University, Villanova, PA 19085, USA}
\author{Pawe\l~Szewczyk}
\affiliation{Astronomical Observatory, University of Warsaw, al. Ujazdowskie 4, 00-478 Warsaw, Poland}
\author{Gabriele~Vedovato}
\affiliation{Universit\`a di Padova, Dipartimento di Fisica e Astronomia, I-35131 Padova, Italy}
\affiliation{INFN, Sezione di Padova, I-35131 Padova, Italy }
\author{Michele~Zanolin}
\affiliation{Embry-Riddle Aeronautical University, Prescott, AZ 86301, USA}

% \collaboration{---}
% \nocollaboration

%%%%%%%%%%%%%%%%%%%%
% LVK opt-in contribution:
\author{Imre~Bartos}
\affiliation{Department of Physics, University of Florida, Gainesville, FL 32611-8440, USA}
\author{Shubhagata~Bhaumik}
\affiliation{Department of Physics, University of Florida, Gainesville, FL 32611-8440, USA}
\author{Tomasz~Bulik}
\affiliation{Astronomical Observatory, University of Warsaw, al. Ujazdowskie 4, 00-478 Warsaw, Poland}
\author{Marco~Drago} \affiliation{Università di Roma La Sapienza, I-00133 Roma, Italy and INFN, Sezione di Roma, I-00133 Roma, Italy}
\author{Jos\'e~A.~Font}
\affiliation{Departamento de Astronom\'{\i}a y Astrof\'{\i}sica, Universitat de Val\`{e}ncia, E-46100 Burjassot, Val\`{e}ncia, Spain}
\affiliation{Observatori Astron\`omic, Universitat de Val\`encia, E-46980 Paterna, Val\`encia, Spain}
\author{Fabio~De~Colle}
\affiliation{Instituto de Ciencias Nucleares, Universidad Nacional Aut\'onoma de M\'exico, A. P. 70-543 04510 D. F. Mexico}
\author{Juan~Garc\'ia-Bellido}
\affiliation{Instituto de F\'isica Te\'orica UAM/CSIC, Universidad Aut\'onoma de Madrid, Cantoblanco 28049 Madrid, Spain}
\author{V.~Gayathri}
\affiliation{Leonard E. Parker Center for Gravitation, Cosmology, and Astrophysics, University of Wisconsin–Milwaukee, Milwaukee, WI 53201, USA}
\affiliation{Department of Physics, University of Florida, Gainesville, FL 32611-8440, USA}
\author{Brennan~Hughey}
\affiliation{Department of Physics and Astronomy, Embry-Riddle Aeronautical University, Prescott, AZ 86301-3720, USA}
\author{Guenakh~Mitselmakher}
\affiliation{Department of Physics, University of Florida, Gainesville, FL 32611-8440, USA}
\author{Tanmaya~Mishra}
\affiliation{Department of Physics, University of Florida, Gainesville, FL 32611-8440, USA}
\author{Soma~Mukherjee}
\affiliation{Department of Physics and Astronomy, University of Texas Rio Grande Valley, Brownsville, TX 78520, USA}
\author{Quynh~Lan~Nguyen}
\affiliation{Department of Physics and Astronomy, University of Notre Dame, 225 Nieuwland Science Hall, Notre Dame, IN 46556, USA}
\author{Man~Leong~Chan}
\affiliation{Department of Physics and Astronomy, University of British Columbia, Vancouver, BC V6T 1Z4, Canada}
\author{Irene~Di~Palma}
\affiliation{Universit\`a di Roma ``La Sapienza'', I-00185 Roma, Italy}
\affiliation{INFN, Sezione di Roma, I-00185 Roma, Italy}
\author{Brandon~J.~Piotrzkowski}
\affiliation{University of Wisconsin-Milwaukee, Milwaukee, WI 53201, USA}
\author{Neha~Singh}
\affiliation{Astronomical Observatory, University of Warsaw, al. Ujazdowskie 4, 00-478 Warsaw, Poland}

% \collaboration{LIGO Scientific Collaboration, Virgo Collaboration}

% \author{Authors}
% \affiliation{Affiliations}

% \collaboration{ASAS-SN Collaboration}

% \author{Authors}
% \affiliation{Affiliations}

% \collaboration{DLT40 Collaboration}

% \author{Other authors}
% \affiliation{Affiliations}

\begin{abstract}

We present the results from a search for gravitational-wave transients associated with core-collapse supernovae observed optically within 30\,Mpc during the third observing run of Advanced LIGO and Advanced Virgo. No gravitational wave associated with a core-collapse supernova has been identified. We then report the detection efficiency for a variety of possible gravitational-wave emissions. For neutrino-driven explosions, the distance at which we reach 50\% detection efficiency is up to 8.9\,kpc, while more energetic magnetorotationally-driven explosions are detectable at larger distances. The distance reaches for selected models of the black hole formation, and quantum chromodynamics phase transition are also provided. We then constrain the core-collapse supernova engine across a wide frequency range from 50\,Hz to 2\,kHz. The upper limits on gravitational-wave energy and luminosity  emission are at low frequencies down to $10^{-4}\,M_\odot c^2$ and $6 \times 10^{-4}\,M_\odot c^2$/s, respectively. The upper limits on the proto-neutron star ellipticity are down to 3 at high frequencies. Finally, by combining the results obtained with the data from the first and second observing runs of LIGO and Virgo, we improve the constraints of the parameter spaces of the extreme emission models. Specifically, the proto-neutron star ellipticities for the long-lasting bar mode model are down to 1 for long emission (1\,s) at high frequency.

\end{abstract}


\maketitle

\date[\relax]{Dated: \today }

%%%%%%%%%%%%%%%%%%%%%%%%%%%%%%%%%%%%%%%%%%%%%%%%%%%%%
\section{Introduction}
\label{sec:introduction}
%%%%%%%%%%%%%%%%%%%%%%%%%%%%%%%%%%%%%%%%%%%%%%%%%%%%%%

The observation of gravitational waves (GWs) from the binary black hole merger in 2015~\cite{Abbott:2016blz} began the field of GW Astronomy. Two years later, a merger of two neutron stars was observed in both GW and electromagnetic spectra~\cite{TheLIGOScientific:2017qsa}. While detecting GWs from compact binary systems has become commonplace~\cite{LIGOScientific:2018mvr,Abbott:2020niy,LIGOScientific:2021djp}, we are waiting for the discovery of GWs from other astrophysical sources. Core-collapse supernovae (CCSNe) are one of them, and the next nearby CCSN will be one of the most interesting astronomical events of the century.

The CCSNe are the violent explosions of massive (above $8\,M_\odot$) stars at the end of their life, giving birth to neutron stars and black holes. While the theoretical understanding of stellar collapse is growing (see recent reviews~\cite{Mezzacappa:2020oyq,Janka:2017vcp,Muller:2020ard,Kotake:2005zn}), the century-old question about the CCSN explosion mechanism~\cite{Janka:2012wk} is not yet solved. The inner dynamics of the CCSN engine can be completely understood with direct observations. Because GWs and neutrinos leave the core around the collapse time and almost do not interact with the star's matter, they are the only probes of the CCSN engine. GWs, in particular, will allow us to measure the engine's dynamics directly. All known CCSNe were observed in the electromagnetic spectrum, and low energy neutrinos were detected so far only from SN~1987A~\cite{hirata:87,bionta:87,1987ESOC...26..237A}. The scientific community awaits a multimessenger observation of neutrinos, EM radiation, and GWs from a nearby CCSN.

Because the GWs predicted from the multidimensional CCSN simulations produce relatively weak GWs (compared to compact binaries), they can be detectable only within Milky Way for Advanced GW detectors and their upgrades~\cite{Szczepanczyk:2021bka,gossan:16}. However, these simulations and the many models of more energetic GW emission have not yet been observationally constrained. This motivates us to search for GW transients with CCSNe that occurred in the nearby Universe. While a Galactic CCSN will be the best opportunity for detecting GWs, LIGO~\cite{TheLIGOScientific:2014jea}, Virgo~\cite{TheVirgo:2014hva}, and GEO600~\cite{Dooley:2015fpa} performed two targeted searches for GWs from CCSNe outside Milky Way. Both of them reported null results. The search with the initial GW detector data~\cite{SNSearchS5A5S6} established the search method; four CCSNe were analyzed within a distance of 15\,Mpc. Later, Ref.~\cite{SNSearchO1O2} reports on a search on five CCSNe within 20\,Mpc with data from the first and second observing runs (O1 and O2). For the first time, the GW data enabled us to exclude parts of the parameter spaces of CCSN extreme emission models. LIGO, Virgo, and KAGRA~\cite{Aso:2013eba} performed all-sky all-time generic searches~\cite{abbott17,PhysRevD.100.024017,KAGRA:2021tnv,LIGOScientific:2022mykPTEP} that also had the potential to detect GWs from a CCSN. The previous CCSN searches are described more in detail in~\cite{Szczepanczyk:2021bka}.

In this paper, we present an optically targeted search with CCSNe observed up to a distance of approximately 30\,Mpc that occurred during the third observing run (O3, 2019 Apr 1 -- 2020 Mar 30) of LIGO, Virgo, and KAGRA. We selected eight CCSNe; seven of them are type-II (SNe 2019ehk, 2019ejj, 2019fcn, 2019hsw, 2020cxd, 2020dpw, 2020fqv) and one is type-Ic (SN 2020oi). No GW associated with a CCSN has been identified, so we provide statements about the predicted GW emissions' detectability and constrain the CCSN engine's dynamics. For the first time, we report the upper limits on the GW luminosity and proto-neutron star (PNS) ellipticity. The constraints of two GW extreme emission models are improved with respect to~\cite{SNSearchO1O2}.

This paper is organized as follows. In Sec.~\ref{sec:targeted_sne}, we list the analyzed CCSNe and outline the method for calculating a period of an expected time of GW emission. Sec.~\ref{sec:methodology} describes the search method and simulated GW signals. Sec.~\ref{sec:results} reports on the results, including the distance reaches for simulated GW signals. Then, Sec.~\ref{sec:constraints} provides constraints on CCSN engine properties. We include generic upper limits of the emitted GW energy, luminosity, and PNS ellipticity. Model exclusion statements are then presented, specifically for the long-lasting bar model. The summary and discussion are in Sec.~\ref{sec:summaries}.

\begin{table*}[!bt]
\caption{
CCSNe selected as optical targets for the GW search described in this paper. The variables $t_{1}$ and $t_{2}$ are the start and end of the on-source windows (OSWs, see Sec.~\ref{sec:osw}), and $\Delta t$ is the OSW durations. The last column shows coincident data durations $T_\mathrm{coinc}$ together with duty factors ($T_\mathrm{coinc} / \Delta t$).\\
}
\centering
\begin{tabular}{lcccllccr}
\hline
\hline
Supernova & Type 
& \multicolumn{1}{c}{Host}
& \multicolumn{1}{c}{Distance}
& \multicolumn{1}{c}{$t_{1}$} & \multicolumn{1}{c}{$t_{2}$}
& \multicolumn{1}{c}{$\Delta t$} & \multicolumn{1}{c}{OSW} 
&\multicolumn{1}{c}{$T_\mathrm{coinc}$}\\
& & \multicolumn{1}{c}{Galaxy} &\multicolumn{1}{c}{[Mpc]} &\multicolumn{1}{c}{[UTC]} 
& \multicolumn{1}{c}{[UTC]} & \multicolumn{1}{c}{[days]} 
& \multicolumn{1}{c}{Method} 
&\multicolumn{1}{c}{[days]}\\
\hline
SN 2019ehk & IIb & NGC 4321 & 16.1 & 2019 Apr 23.10 \,\,  & 2019 Apr 24.50 \,\, & 1.40 & 2 & 0.41 (29\%) \\
SN 2019ejj & II & ESO 430-G20 & 15.7 & 2019 Apr 23.28 \,\, & 2019 Apr 30.86 \,\, & 7.58 & 3 & 1.25 (16\%) \\
SN 2019fcn & II & ESO 430-G20 & 15.7 & 2019 May 03.02 \,\, & 2019 May 07.56 \,\, & 4.54 & 3 & 2.51 (55\%) \\
SN 2019hsw & II & NGC 2805 & 28.2 & 2019 Jun 05.14 \,\, & 2019 Jun 13.14 \,\, & 8.00 & 1 & 5.08 (64\%) \\
SN 2020oi  & Ic & NGC 4321 & 16.1 & 2020 Jan 02.48 \,\, & 2020 Jan 06.18 \,\, & 3.70 & 2 & 2.56 (69\%) \\
SN 2020cxd & IIP & NGC 6395 & 20.9 & 2020 Feb 16.53 \,\, & 2020 Feb 22.53 \,\, & 6.00 & 1 & 4.58 (76\%) \\
SN 2020dpw & IIP & NGC 6952 & 22.3 & 2020 Feb 21.08 \,\,& 2020 Feb 25.08 \,\, & 4.00 & 1 & 3.06 (77\%) \\
SN 2020fqv & IIb & NGC 4568 & 17.3 & 2020 Mar 22.00 \,\, & 2020 Mar 28.00 \,\, & 6.00 & 1 & 4.06 (68\%) \\
\hline
\hline
\end{tabular} 
\label{tab:sne}
\end{table*}

%%%%%%%%%%%%%%%%%%%%%%%%%%%%%%%%%%%%%%%%%%%%%%%%%%%%%
\section{Targeted Core-Collapse Supernovae} 
\label{sec:targeted_sne} 
%%%%%%%%%%%%%%%%%%%%%%%%%%%%%%%%%%%%%%%%%%%%%%%%%%%%%

\begin{figure}[b]
\centering
\includegraphics[width=1.00\columnwidth]{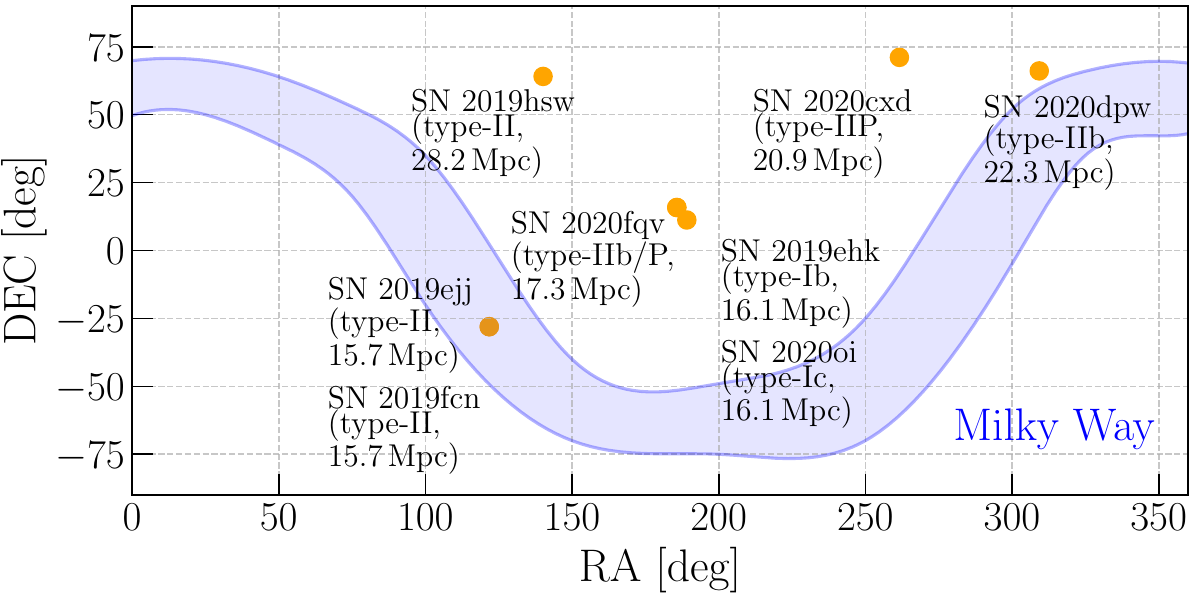}
\caption{
Sky locations of CCSNe analyzed in this paper. 
All were recorded within 30\,Mpc during the third observing run of LIGO, Virgo, and KAGRA.
}
\label{fig:sne}
\end{figure}

%%%%%%%%%%%%%%%%%%%%%%%%%%%%%%%%%%%%%%%%%%%%%%%%%%%%%
\subsection{Source selection}
\label{sec:selection} 
%%%%%%%%%%%%%%%%%%%%%%%%%%%%%%%%%%%%%%%%%%%%%%%%%%%%%

From all CCSNe observed during O3, we have selected those that meet the following criteria: 
(i) they contribute to the model exclusion statements (see Sec.~\ref{sec:mep}), their distances are less than approximately 30\,Mpc,
(ii) the period where we expect to find a GW transient is sufficiently well identified (up to a few day, see Sec.~\ref{sec:osw}); and
(iii) there is sufficient GW detector data coverage to allow us to accumulate a few years of background data (see Sec.~\ref{sec:methodology}). During O3, astronomers found and followed-up numerous CCSNe in the nearby Universe. Based on the information from Astronomical Telegrams~\cite{Rutledge1998} and supernova catalogs (ASAS-SN~\cite{Shappee2014, 10.1093/mnras/stz073, 10.1093/mnras/stx1544, 10.1093/mnras/stx057}, DLT40~\cite{2017ATel10638....1T}, Gaia~\cite{2016A&A...595A...1G,2016A&A...595A...2G}, ASRAS~\cite{www:asras}, TNS~\cite{2018ApJ...853...62T}, OSC~\cite{2017ApJ...835...64G}, CBAT~\cite{www:cbat}), we found eight supernovae of interest. They are reported in Table~\ref{tab:sne} and Figure~\ref{fig:sne} shows their sky locations. Most of them are type-II supernovae originating from red supergiant progenitor stars, just one is type-I. The host galaxies are identified for all of them. The distance to each CCSNe is determined using the estimated distance to its host galaxy.

\textit{SN~2019ehk}, a type-IIb supernova, was discovered on 2019 April 29 22:27:50 UTC~\cite{2019TNSTR.666....1G}. The host galaxy is NGC~4321 (M100) at a distance of 16.1\,Mpc~\cite{Nakaoka:2020rxm, YoungSupernovaExperiment:2021fur, Yao:2020ley} (distance inferred from Cepheid observations~\cite{De:2020gwx, Jacobson-Galan:2020zxn, Jacobson-Galan:2020gyk}). The progenitor star is either a zero-age main sequence (ZAMS) star with a mass of around $9-9.5\,M_\odot$. The mass-loss rate smaller than $10^{-5}\,M_\odot$/s for a wind velocity 500\,km/s at distances $10^{16}-10^{17}$\,cm from the exploding core~\cite{Jacobson-Galan:2020gyk}.

\textit{SN~2019ejj}, a type-II supernova, was discovered on
2019 May 02 06:18:43 UTC~\cite{2019TNSTR.687....1T}. The host
galaxy is ESO~430-G20 at a distance of 15.7\,Mpc~\cite{Chang_2021} (distance inferred from Tully-Fisher method ~\cite{2007A&A...465...71T}).

\textit{SN~2019fcn}, a type-II supernova, was discovered on
2019 May 08 23:02:24 UTC~\cite{2019TNSTR.766....1S}. The host galaxy is ESO~430-G20 at a distance of 15.7\,Mpc (distance inferred from Tully-Fisher method ~\cite{2007A&A...465...71T}).

\textit{SN~2019hsw}, a type-II supernova, was discovered on 2019 June 18 03:07:12 UTC~\cite{2019TNSTR1030....1S}. The host galaxy is NGC~2805 at a distance of 28.2\,Mpc~\cite{De:2019xhw} (distance inferred from Tully-Fisher method ~\cite{2007A&A...465...71T}).

\textit{SN~2020oi}, a type-Ic supernova, was discovered on
2020 January 07 13:00:54 UTC~\cite{2020TNSTR..67....1F}. The host galaxy is NGC 4321 at a distance of 16.1\,Mpc \cite{Maeda:2021jjl, YoungSupernovaExperiment:2021fur, Tinyanont:2021vkg, Rho:2021} (distance inferred from Cepheid observations \cite{De:2020gwx, Jacobson-Galan:2020zxn, Jacobson-Galan:2020gyk}) . The mass-loss rate is around $1.4 \times 10^{-4}\,M_\odot\,\mathrm{s}^{-1}$ for a wind velocity $100\,\mathrm{km\,s}^{-1}$. SN~2020oi was also observed by \textit{Swift}~\cite{Burrows:2005gfa} with X-ray telescope in the energy range from 0.3\,keV to 10\,keV~\cite{Horesh:2020kwa}.

\textit{SN~2020cxd}, a type-II supernova, was discovered on
2020 February 19 12:44:08 UTC~\cite{2020TNSTR.555....1N}. The host
galaxy is NGC~6395 at a distance of 20.9\,Mpc \cite{Kozyreva:2022wkp, Yang:2021fka, Perley:2020ajb} (distance inferred from Tully-Fisher method \cite{Valerin:2022fxg}). The progenitor star has a ZAMS mass of $\lesssim 15\,M_\odot$.

\textit{SN~2020dpw}, a type-IIP supernova, was discovered on
2020 February 26 10:01:22 UTC~\cite{2020TNSTR.653....1W}. The host
galaxy is NGC~6952 at a distance of 22.3\,Mpc (distance inferred from Tully-Fisher and SN type-Ia methods, both agree)~\cite{Chang_2021}.

\textit{SN~2020fqv}, a type-IIb supernova, was discovered on
2020 March 31 08:06:02 UTC~\cite{2020TNSTR.914....1F}. The host galaxy is NGC~4568 at a distance of 17.3\,Mpc (distance inferred from Tully-Fisher method \cite{2007A&A...465...71T}). The progenitor star has a ZAMS mass of $13.5 - 15\,M_\odot$~\cite{Tinyanont:2021cwl}.

\begin{figure*}[t] 
\centering 
\includegraphics[width=0.90\textwidth]{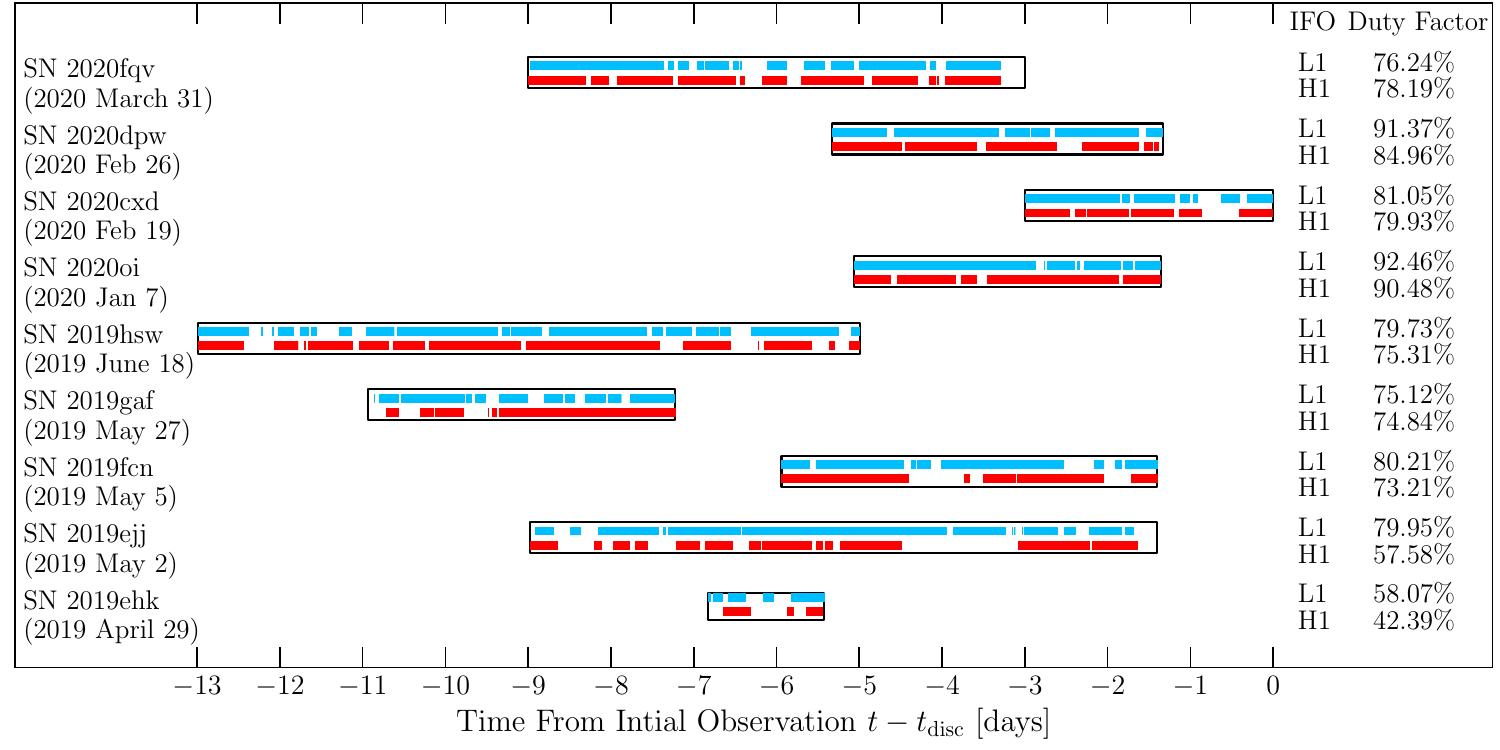} 
\caption{
Visual representation of the on-source windows (see Sec.~\ref{sec:osw}), the data coverage for each detector, and the detector duty factors (percentage of available data inside the on-source window). These windows are plotted with respect to the discovery time $t_\mathrm{disc}$, and the brackets show CCSN discovery dates in UTC. The plotted interferometers (IFO) are LIGO Hanford (H1) and LIGO Livingston (L1).
}
\label{fig:coverage}
\end{figure*} 
 
%%%%%%%%%%%%%%%%%%%%%%%%%%%%%%%%%%%%%%%%%%%%%%%%%%%%%%%%%%
\subsection{On-source window}
\label{sec:osw}
%%%%%%%%%%%%%%%%%%%%%%%%%%%%%%%%%%%%%%%%%%%%%%%%%%%%%%%%%%

The on-source window (OSW) is a time interval where we search for the GW transient. It is defined as $[t_1, t_2]$, where $t_1$ and $t_2$ are the beginning and end times, respectively. The methodology to compute $t_1$ and $t_2$ is a compromise between achieving a high probability (see~\cite{Gill:2022amm} for reference of some of the limitations in achieving 100\% probability) to contain the possible GW emission from the CCSN and, at the same time, do not lose the benefits of having a shorter time window than the all-sky searches, e.g.~\cite{KAGRA:2021tnv}. See Section~\ref{sec:cwb} for a detailed explanation of these benefits.

While we currently do not know how to quantify all the uncertainties of the OSW calculation process exactly (for example, uncertainties for the values taken from~\cite{Barker:2021iyr} or different idiosyncracies of telescopes sensitivities), we aimed for a $2\sigma$ probability for the OSW to contain the GW emission. We use three different methods to calculate OSWs. The choice depends on the availability of early observation (pre-peak luminosity) data and the availability of published tools to model the light curves. Table~\ref{tab:sne} shows the calculated OSWs and methods used.

\subsubsection{Method 1 - The Early Observation}

For SNe~2019hsw, 2020cxd, 2020dpw, and 2020fqv, the available luminosity public data, at the time of this analysis, was of poor quality and did not appear to include the peak luminosity. For these candidates, we use an \textit{Early} method described in~\cite{SNSearchO1O2} with an increased upper limit on the delay between the collapse of the shock breakout as described below. We note that a detailed photometry~\cite{Tinyanont:2021cwl} for SN~2020fqv was publicly available in the late stages of our analysis.

In this method, the initial supernova observation (discovery time, $t_\mathrm{disc}$) is used as the end of the OSW, $t_2 = t_\mathrm{disc}$\footnote{While we show in Figure~\ref{fig:coverage} the calculated OSWs, due to a bookkeeping error, three extra days after the discovery time were included in the analysis of SN~2020cxd. However, it has a marginal effect on the search results.}. The beginning of the OSW is calculated using the last null observation of a CCSN in the host galaxy, $t_\mathrm{Null}$. This time can still be after the GW emission since the shock breakout (SBO) is delayed with respect to the collapse, $\Delta t_\mathrm{SBO}$. We use $t_1 = t_\mathrm{Null} - \Delta t_\mathrm{SBO}$ to account for this. Ref.~\cite{Barker:2021iyr} derives a relationship between the progenitor star mass and $\Delta t_\mathrm{SBO}$. Furthermore, an extensive study was performed in~\cite{Smartt:09} about the mass distribution of the CCSNe progenitor masses in the local universe (within 30\,Mpc where this study is feasible). As summarized in Fig.~6 of Ref.~\cite{Smartt:09}, there is an apparent upper limit of CCSNe progenitor masses at 16.5 solar masses in the local universe. In the absence of progenitor mass information for the four CCSN candidates, the mass delay relationship in Ref.~\cite{Barker:2021iyr} indicates $\Delta t_\mathrm{SBO} = 3$ days as an appropriate upper limit for the collapse SBO delay. It is worth mentioning that the last null time estimation from the telescope depends on the relative sensitivity to the one at the discovery time~\cite{Gill:2022amm}.

\subsubsection{Method 2 - The Quartic Interpolation}

\begin{figure}[ht]
\centering
\includegraphics[width=1.00\columnwidth]{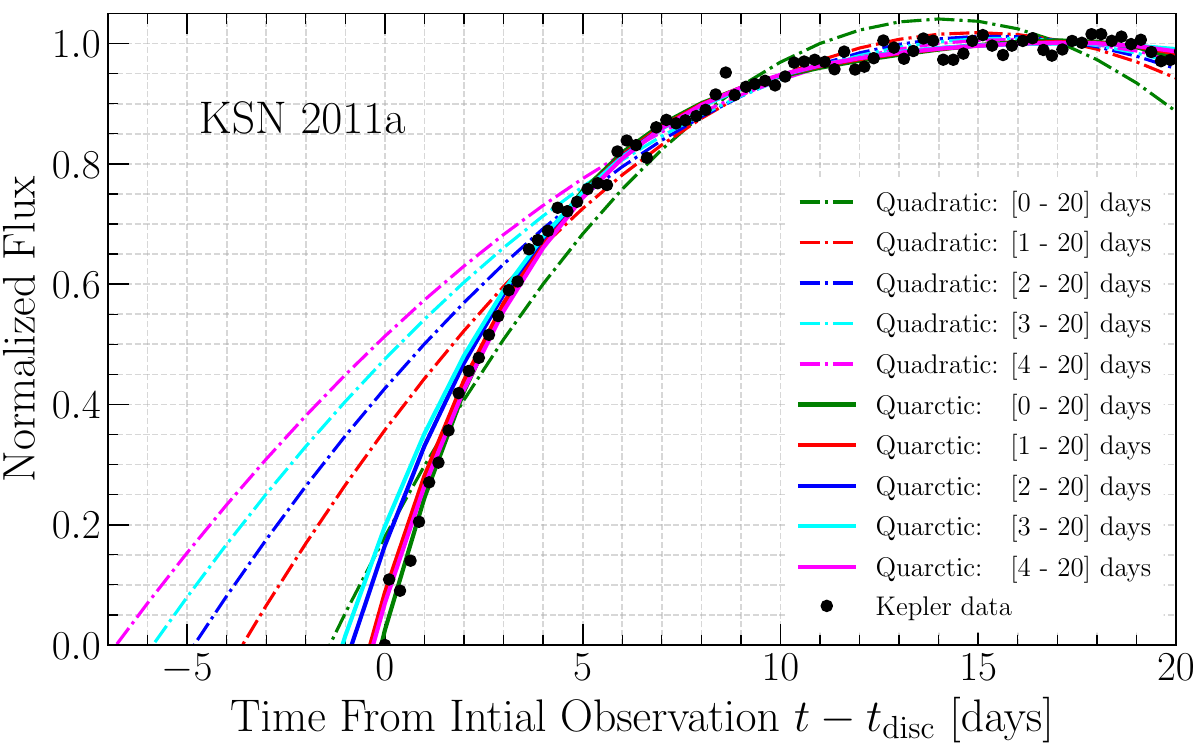}
\caption{The shock breakout estimation using quadratic ($2^\mathrm{nd}$ degree) and quartic ($4^\mathrm{th}$ degree) polynomials. Because shock breakout was observed for KSN~2011a~\citep{Garnavich:2016thk,astropy:2013, astropy:2018, astropy:2022}, it allows testing usage of the polynomial interpolations in a case when a CCSN is discovered up to a few days after the shock breakout. While a quartic fit is reliable, the quadratic fit introduces biases.}
\label{fig:method2}
\end{figure}

For SNe~2019ehk and 2020oi, the light curve indicated that the discovery happened in the homologous phase of the shock wave expansion. Since the part of the light curves of CCSNe up to the flux peak has a high degree of universality among observed CCSNe, we investigated what would be the simplest polynomial fit (where we estimated the coefficients of different polynomials with a chi-square minimization between the data and the polynomials). We used, as a testing ground, a detailed light curve (since the moment of observed SBO) for KSN~2011a provided by Kepler~\cite{Garnavich:2016thk} and Astropy~\citep{astropy:2013, astropy:2018, astropy:2022}. Figure~\ref{fig:method2} shows possible interpolations in cases when a potential first observation can be up to a few days after the SBO. A simple quadratic ($2^\mathrm{nd}$ degree) polynomial fit reveals biases. The estimated time of the SBO can be earlier than the real SBO time because the slope of the light curve around the peak is smaller than in the early stages of the light curve. Next, we tested a quartic ($4^\mathrm{th}$ degree) polynomial fit that more accurately estimates SBO time even when the first few days of data are missing. Therefore, we adopted a quartic polynomial fit.

\begin{figure*}[hbt] 
  \begin{minipage}[c][][t]{0.495\textwidth}
    \vspace*{\fill}
    \flushleft
    \includegraphics[width=0.96\linewidth]{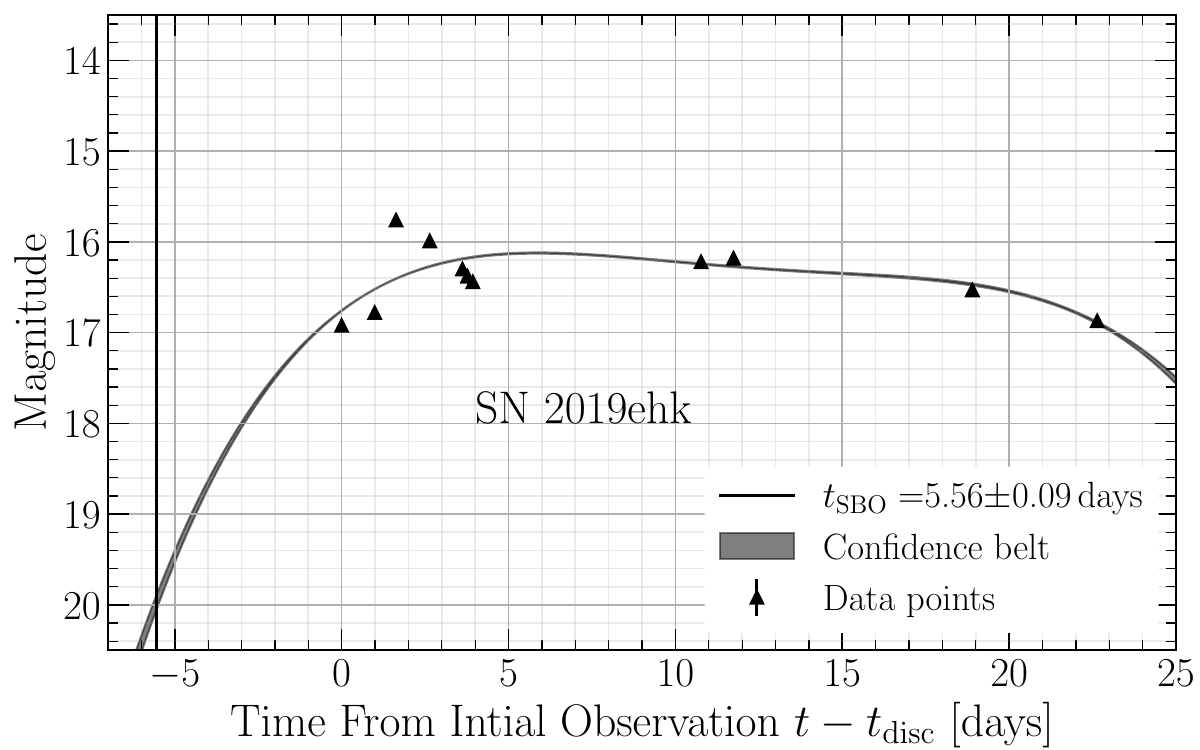}
  \end{minipage}%
  \begin{minipage}[c][][t]{0.495\textwidth}
    \vspace*{\fill}
    \flushright
    \includegraphics[width=0.96\linewidth]{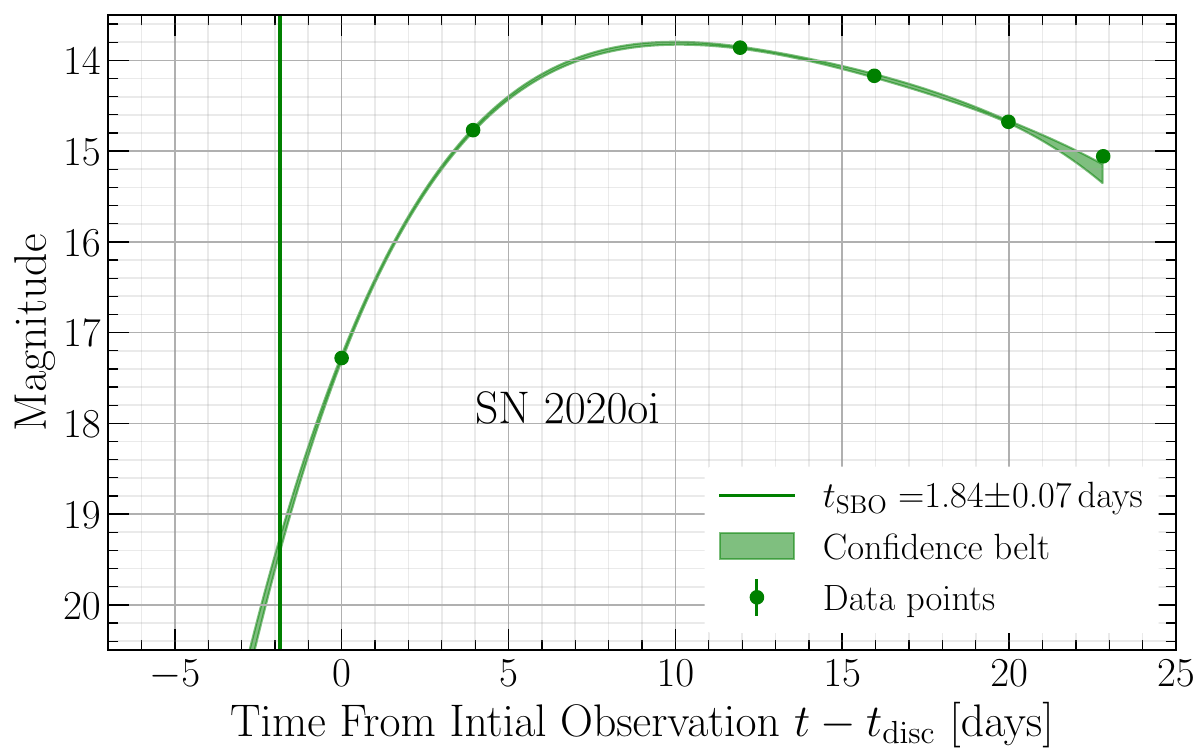}
  \end{minipage}
  \caption{
The estimation of the shock breakout, $t_\mathrm{SBO}$, for SNe~2019ehk and 2020oi was performed with the quartic interpolation. The r-band magnitudes, including the pre-shock breakout values, are public and taken from Refs.~\cite{Jacobson-Galan:2020gyk,Smith_2019,ZTF19aatesgp,ZTF20aaelulu}. The quartic polynomial coefficients are determined with a chi-square minimization. The estimated time of the SBO (marked by the vertical lines) is determined by the intersection of the curves with the average pre-SBO values. The uncertainty in the time of the SBO is determined from the standard deviation of the estimated SBO times if we perform the interpolations for the data randomized within the telescope uncertainties provided for each data point. 
}
  \label{fig:quartic_sne}
\end{figure*}

Given that the measured r-band magnitudes~\cite{Jacobson-Galan:2020gyk,Smith_2019,ZTF19aatesgp,ZTF20aaelulu} at any time have uncertainties, we performed Monte-Carlo simulations where individual values of the optical flux were randomly varied within the given uncertainty from the telescope. In each repetition, the least square estimate of the time of the SBO was performed. As a result, for each candidate, we obtained confidence belts of randomized estimates of the SBO time, shown in Figure~\ref{fig:quartic_sne}. We associate an uncertainty equal to $2\sigma$ of this confidence belt at the estimated time of the shock breakout. We then obtained for each candidate a histogram of randomized estimates of the SBO. In order to take into account the uncertainties in the estimates of the pre-shock breakout flux of the value, we added 0.5~days to the estimated shock break-out time as the double standard deviation. This value came from studying the test Kepler data, the systematics of the conversion between normalized flux and  magnitude for these candidates, and we also studied the systematics of these interpolation processes in other CCSNe for which open data is available even when no GW observing run was undergoing.

Using the methodology described above, we estimated the $t_\mathrm{SBO}$ for each of the candidates and the uncertainties and, from that, the estimates of $t_1$ and $t_2$. The vertical lines in Figure~\ref{fig:quartic_sne} show $t_\mathrm{SBO}$ which is determined by the intersection of the quartic fitted curves with the pre-SBO average magnitude. SN~2019ehk also had a known progenitor mass of about $9.5 \, M_\odot$, which according to Ref.~\cite{Barker:2021iyr} corresponds to a $1.5 \pm 0.2$~days delay between the collapse and the SBO. In this case, $t_1$ became $t_1 = t_\mathrm{disc} - t_\mathrm{SBO} - \Delta t_\mathrm{delay} - \Delta t_\mathrm{SBO,1}$ where the $\Delta t_\mathrm{delay}=1.5$~days is the estimated delay of the SBO with respect to the collapse, and $\Delta t_\mathrm{SBO,1}$ conservatively estimated at $ 0.7$~days account for the joint uncertainty in the $t_\mathrm{SBO}$ estimate and the estimate of $\Delta t_\mathrm{delay}$. For the estimate of $t_1$ in SN~2020oi, since  the  mass  of  the progenitor was unknown we used $t_1 = t_\mathrm{disc} - t_\mathrm{SBO} - \Delta t_\mathrm{SBO,1}$ choosing $\Delta t_\mathrm{SBO} = 4.5$~days, to include in a conservative way the 3~days upper limit on the collapse to break out as described in the conservative method section (based on Ref.~\cite{Barker:2021iyr}) and the error in the estimate of the time of the shock break out. The end of the OSWs, $t_2$, were determined using $\Delta t_\mathrm{SBO,2}$ where $t_2 = t_\mathrm{disc} - t_\mathrm{SBO} - \Delta t_\mathrm{delay} + \Delta t_\mathrm{SBO,2}$. For each of the two candidates, $\Delta t_\mathrm{SBO,2}$ was then calculated by adding both the uncertainty of the delay with the double standard deviation from the quartic interpolation, choosing conservatively $\Delta t_\mathrm{SBO,2}=0.7$~days. $\Delta t_\mathrm{delay}$ was chosen as 1.5 days, the minimum collapse to shock breakout delay in Figure 6 of Ref.~\cite{Barker:2021iyr}.

\subsubsection{Method 3 - The Physics-Based}

For SNe 2019fcn and 2019ejj, good quality public data were unavailable to perform the quartic interpolations. Instead, we used the OSWs published in Ref.~\cite{Gill:2022amm} based on Las Coumbres Observatory data (named there an~\textit{Updated EOM}). These OSWs were produced using the overlap of different physics-based interpolation methods, with a large probability of containing the GW emission. This method has the potential also to provide progenitor star properties. While Methods~1 and~2 are frequentists, the results in this method involve Bayesian posteriors.

%%%%%%%%%%%%%%%%%%%%%%%%%%%%%%%%%%%%%%%%%%%%%%%%%%%%%
\section{Methodology}
\label{sec:methodology}
%%%%%%%%%%%%%%%%%%%%%%%%%%%%%%%%%%%%%%%%%%%%%%%%%%%%%%%%

We closely follow the methodology presented in Ref.~\cite{SNSearchO1O2}, and this section briefly summarizes the method. The search is performed at the frequency range from 16\,Hz to 2048\,Hz.

\subsection{Data}

The LIGO and Virgo detectors use a photon recoil-based calibration~\cite{Karki:2016pht,Cahillane:2017vkb,Viets:2017yvy} resulting in a complex-valued, frequency-dependent detector response. Previous studies have documented the systematic error and uncertainty bounds for O3 strain calibration in LIGO~\cite{Sun_2020,Sun2021} and Virgo~\cite{Acernese2018, Acernese2021}. The product of this calibration is strain data sampled at 16384\,Hz. Times affected by transient noise sources, referred to as~\textit{glitches}, and other data quality issues are identified so that searches for GWs can exclude (veto) these periods of poor data quality~\cite{TheLIGOScientific:2016zmo,LIGOScientific:2019hgc,Davis:2021ecd,nguyen2021environmental,Fiori2020,Virgo:2022kwz,Virgo:2022ysc}. In addition, several known persistent noise sources are subtracted from the data using information from witness auxiliary sensors~\cite{Driggers:2018gii,Davis:2018yrz}. In this search, we analyze a network of LIGO detectors (Hanford and Livingston). While adding the Virgo detector may provide benefits, e.g. sky localizations, it can affect the confidence of candidate detections. Hence, it is not an optimal choice for detection purposes~\cite{KAGRA:2021tnv} because of the higher rate of glitches.

\subsection{Coherent WaveBurst}
\label{sec:cwb}

We use coherent WaveBurst (cWB), a model-independent search algorithm for detecting and reconstructing GWs~\cite{cwb}. The cWB searches for a coherent signal power in multiple detectors. The analysis is performed in a wavelet domain on GW strain data using the multi-resolution Wilson-Daubechies-Meyer wavelet transform~\cite{Necula:2012zz}. The algorithm selects wavelets with amplitudes above the fluctuations of the detector noise and groups them into clusters, identifying coherent events.

The events are ranked with $\eta_{\rm c}=\sqrt{E_{\rm c}/\max(\chi^2,1)}$, where $E_{\rm c}$ is coherent energy, and $\chi^2$ quantifies the agreement of cWB reconstruction and the detector data. The correlation coefficient $c_{\rm c} = E_{\rm c}/(E_{\rm c}+E_{\rm n})$, where $E_{\rm n}$ is a residual energy, further reduces noise events. The events are accepted when $\chi^2<2.5$ and $c_{\rm c}>0.8$. The remaining ones are divided into two mutually exclusive classes. Class~\textit{C1} contains transients of a few cycles with a prime example of short (order of 10\,ms) and broadband (order of 100\,Hz) \textit{blip} glitches~\cite{abbott:16a-detchar,Cabero:19}. All other events are in class~\textit{C2}.

As usual, we use time-shifting analysis to estimate the background for the burst searches. This method allows for estimating the false-alarm rate (FAR). For each analyzed OSW, we accumulated a few years of background data. An event with the smallest FAR from the non-time-shifted analysis across the two search classes is called the \textit{loudest event} and is considered a potential GW candidate. Given two search classes, a trial factor of 2 is applied to the event's FAR~\cite{SNSearchO1O2}. The loudest event significance is assessed by calculating its false-alarm probability (FAP). This is the probability of obtaining one or more noise events that are less than or equally ranked:
\begin{equation}
  \label{eqn:fap}
  \mathrm{FAP} = 1-\exp{(-T_\mathrm{coinc} \times \mathrm{FAR})} \,
\end{equation}
where $T_\mathrm{coinc}$ is the coincident data duration of the appropriate OSW.

Following Sec.~\ref{sec:osw}, reducing time intervals to search for GW emission benefits optically targeted searches compared to all-sky searches in two ways or a combination. The first is to increase the statistical significance of the GW candidates by reducing the event's FAR. The second is to achieve the same statistical significance but with a larger FAR. This second option, in turn, allows reconstructing events with a smaller $\eta_{\rm c}$ or GWs emitted further away from the Earth. The usefulness of this second approach depends on the slope of the FAR vs $\eta_{\rm c}$ histograms (a shallower slope, like in the case of glitchy data, would give larger gains in the detection range). Hybridization of the two would mean using some benefits in both directions. In this paper, we only use the first type of benefit. 

Note that the previous searches~\cite{SNSearchO1O2,SNSearchS5A5S6} used an estimated 9.1\% for the calibration error. This conservative value is revisited in Appendix~\ref{sec:cal}. In brief, the predicted GW signals from CCSNe (Sec.~\ref{sec:wave}) were distorted in time and frequency according to the detector's calibration errors at the times of the OSWs. We find that the impact on the $\eta_{\rm c}$ and $c_{\rm c}$ is negligible for a network of LIGO detectors.

%%%%%%%%%%%%%%%%%%%%%%%%%%%%%%%%%%%%%%%%%%%%%%%%%%
\subsection{Search Sensitivity}
\label{sec:wave}
%%%%%%%%%%%%%%%%%%%%%%%%%%%%%%%%%%%%%%%%%%%%%%%%%%

The search sensitivity is determined by adding waveforms to the detector strain data around every 33\,s and reconstructing them with cWB. The procedure is repeated for sources placed at a range of distances, constructing a \textit{detection efficiency}. The simulated sources are placed at the sky positions of the analyzed CCSNe. We measure the search sensitivity with waveforms calculated from multidimensional CCSN simulations: neutrino- and magnetohydrodynamically-driven explosions, black hole formation, and quantum chromodynamics phase transition. We also consider two extreme emission models. For each three-dimensional model, the source orientation is randomized, and the waveforms from two-dimensional CCSN simulations are averaged over the source orientation using a factor of $\sqrt{5/18}$~\cite{SNSearchO1O2}. The distance at 50\% detection efficiency is called the \textit{detection reach}.

As mentioned earlier, we employ \textit{ad-hoc} signals to estimate the search sensitivity to monochromatic CCSN GW emission. These signals do not have physical meaning but are used to constrain the properties of a CCSN engine at a given frequency bin. The detection efficiencies are calculated as a function of the signal's root-sum-squared strain:
\begin{equation}
\label{eqn:hrss}
\hrss=\sqrt{\int \left( h^2_+(t) + h^2_\times(t) \right) \mathrm{d}t}\,,
\end{equation}
where $h_+$ and $h_\times$ are GW polarizations.

\subsubsection{Neutrino-driven explosions}
\label{sec:NeutrinoDrivenExplosions}

In this section, we list and briefly describe the non-rotating, neutrino-driven explosion, or non-exploding CCSN models that are used to test the sensitivity of the search. All of the waveforms were simulated in three dimensions.

The first is model s11 from {\it Andresen et al. 2017}~\cite{Andresen2017} (And+17). It is a solar metallicity $11.2\,M_\odot$ progenitor star. The shock is not revived in this model. The simulation ends 352\,ms after the core bounce. No standing accretion shock instability (SASI) is observed and most of the GW amplitudes are produced by convection. The low mass and lack of shock revival result in low GW amplitudes, with the majority of the GW energy at frequencies between 500\,Hz and 700\,Hz. 

The s15 model is the solar metallicity, $15\,M_\odot$ progenitor star simulated by {\it Kuroda et al. 2016}~\cite{kuroda2016} (Kur+16) with the SFHx equation of state (EoS). The GW signal reaches frequencies of up to 1000\,Hz, and has a strong low-frequency component due to the SASI. The simulation was stopped $\sim350$\,ms after the core bounce before the shock revival.

The C15 model is another $15\,M_\odot$ progenitor star simulated by {\it Mezzacappa et al. 2020}~\cite{mezzacappa2020} (Mez+20). The GW signals show a low-frequency component due to the SASI, and also a high-frequency component above 600\,Hz due to convection. The simulation was stopped 450\,ms after the core bounce before shock revival. The signal reaches very high GW amplitudes of above 1500\,Hz. 

The L15-3 model by {\it M\"uller et al. 2012}~\cite{mueller:e12} (Mul+12) is an older simulation. However, we still include this model to compare to results in previous searches directly. The progenitor is a $15\,M_\odot$ star. The GW emission only reaches frequencies of $\sim 500$\,Hz, which is much lower than more modern simulations. The GW signal is 1.4\,s long, and the model successfully undergoes shock revival.  

We use two models from {\it O'Connor and Couch 2018}~\cite{oconnor2018} (Oco+18). They are both simulations of a $20\,M_\odot$ progenitor star from MESA~\cite{Paxton2011, Paxton2013, Paxton2015, Paxton2018, Paxton2019, Jermyn2023}. The first model, m20, does not include perturbations, and the second model, m20p, has perturbations. Both of them are simulated for over 500\,ms, and do not undergo shock revival before the end of the simulations. The GW signals in both models have a low-frequency component due to the SASI, and the high-frequency modes reach frequencies of over 1000\,Hz.

The next two signals were simulated by {\it Powell and M\"uller 2019}~\cite{powell2019} (Pow+19). The he3.5 model is an ultra-stripped $3.5\,M_\odot$ helium progenitor star. The low mass results in low GW amplitudes. The s18 model is a solar metallicity $18\,M_\odot$ progenitor star. Both models undergo successful neutrino-driven explosions and therefore have no low-frequency SASI modes. The he3.5 model is 0.7\,s long, and s18 is 0.89\,s. Both models have the majority of their GW energy between 600\,Hz and 1000\,Hz. 

We use three models from {\it Radice et al. 2019}~\cite{radice2019} (Rad+19). They are all solar metallicity progenitor stars with masses of $9\,M_\odot$ model s9, $13\,M_\odot$ model s13, $25\,M_\odot$ model s25. The s9 simulation undergoes shock revival shortly after the bounce, and model s25 undergoes shock revival much later in the simulation, s13 does not explode. The s25 model has low-frequency GW emission due to the SASI, the GW signal is $\sim 600$\,ms long and reaches frequencies of up to 1200\,Hz. The s9 model has low GW amplitudes due to low mass, is simulated for $\sim 350$\,ms and reaches frequencies of $\sim 1000$\,Hz. The s13 model was simulated for over 0.7 s, and reached frequencies of up to $\sim 2000$\,Hz.

%%%%%%%%%%%%%%%%%%%%%%%%%%%%%%%%%%%%%%%%%%%%%%%%
\subsubsection{Magnetorotational explosions}
%%%%%%%%%%%%%%%%%%%%%%%%%%%%%%%%%%%%%%%%%%%%%%%%

{\it Obergaulinger et al. 2020}~\cite{Obergaulinger:2021} (Obe+20) performed a series of three-dimensional simulations of magnetorotational core collapse of potential progenitors of long gamma-ray bursts. While all of their cores are based on a star with a ZAMS mass of $35 \, M_{\odot}$ and sub-solar metallicity, the authors varied the pre-collapse magnetic field strength and geometry. We select their model signal\_O, which develops a delayed, moderately energetic explosion dominated by magnetically driven jets reaching outflow speeds of around $c/3$ by the end of the simulation at about $0.8$\,s after the bounce. The explosion is launched by a combination of neutrino heating, rotation, the magnetic field, and hydrodynamic instabilities, among which convection is more important than the SASI. Due to greater core asymmetries during the collapse, the GW emissions are larger than for neutrino-driven explosions.

%%%%%%%%%%%%%%%%%%%%%%%%%%%%%%%%%%%%%%%%%%%%%%%%
\subsubsection{Black Hole Formation}
%%%%%%%%%%%%%%%%%%%%%%%%%%%%%%%%%%%%%%%%%%%%%%%%

The black hole formation s40 model is a $40\,M_\odot$ progenitor star, labeled model NR, from {\it Pan et al. 2021}~\cite{pan2021} (Pan+21). The simulation did not undergo shock revival, and it was stopped $\sim 0.7$\,s after the bounce when a black hole was formed. The GW frequency is rather high, above $2000$\,Hz at peak luminosity.

%%%%%%%%%%%%%%%%%%%%%%%%%%%%%%%%%%%%%%%%%%%%%%%%%
\subsubsection{QCD Phase Transition}
%%%%%%%%%%%%%%%%%%%%%%%%%%%%%%%%%%%%%%%%%%%%%%%%%

{\it Kuroda et al. 2022} \cite{Kuroda:2021eiv} (Kur+22) investigate observable signatures of a first-order quantum chromodynamics phase transition in the context of CCSNe. The authors conduct axially symmetric numerical relativity simulations using a hadron–quark hybrid EoS with multi-energy neutrino transport. The progenitor model with $50\,M_\odot$ of Umeda \& Nomoto~\cite{Umeda:2007wk} is a blue supergiant star with solar metallicity. During the post-second-bounce phase, the GWs show a high-frequency emission at the range of $\gtrsim$1\,kHz. We analyzed the waveform from the s50 model.

%%%%%%%%%%%%%%%%%%%%%%%%%%%%%%%%%%%%%%%%%%%%%%%%%
\subsubsection{Extreme Supernova Models}
\label{sec:extreme}
%%%%%%%%%%%%%%%%%%%%%%%%%%%%%%%%%%%%%%%%%%%%%%%%%

Given the large uncertainties in the numerical modeling of fast-rotating CCSNe, it is worth exploring extreme models at the limit of what is possible during the collapse of massive stars. In particular, we consider two scenarios used in previous targeted searches \citep{SNSearchS5A5S6}.

In the \textit{Long-Lasting Bar Mode} scenario (or shortly bar model), a very rapidly rotating progenitor star induces bar mode instabilities in the proto-neutron star. These instabilities are either of dynamical  \cite{lai:95,brown:01,shibata:05,rmr:98} or corrotational (low $T/|W|$) type \cite{rotinst:05,ott:07prl, scheidegger:10b,takiwaki:16,Shibagaki:20,pan2021,Bugli:22}, the latter being preferred according to the most recent and sophisticated simulations. In this scenario, the instability leads to large amplitude GWs that depend on the properties of the deformed proto-neutron star. We use a simple phenomenological bar model~\cite{ott:10dcc,Cerda-Duran:21} in which the waveform can be modeled effectively by sine-Gaussian waveforms with amplitude $h_0$, frequency $f_0$ and width $\tau$ as parameters:
\begin{eqnarray}
h_+(t) &=& h_0 \frac{1 + \cos ^ 2 \iota}{2} \mathrm{e}^{-t^2 / \tau^2} \cos (2 \pi f_0 t), \label{eqn:sg1} \\
h_\times (t) &=& h_0 \cos \iota \,\mathrm{e}^{-t^2 / \tau^2} \sin(2 \pi f_0 t)\,,
\label{eqn:sg2}
\end{eqnarray}
where $\iota$ is a source viewing angle. The model used in the previous targeted searches~\citep{SNSearchO1O2,SNSearchS5A5S6} assumed a cylindrical bar's particular shape, parametrized by its mass, radius, and length. Here, the model is generalized, and no assumption is made about the core shape. Its shape is characterized by frequency, ellipticity, and quadrupolar mass moment. The detailed derivation can be found in Ref.~\cite{Cerda-Duran:21}, and here we provide an overview of a method.

In quadrupolar approximation, a GW signal in the transverse-traceless (TT) gauge is defined by \citep[see e.g.][]{Misner:1973prb}:
\begin{equation}
    h_{ij}^{TT}(t,\mathbf{x}) = \frac{2}{D} \frac{G}{c^4} P_{ij}^{kl} \Ddot{I_{kl}}(t-D/c,\mathbf{x}),
\end{equation}
where $P_{ij}^{kl}$ is the $TT$ projector operator, $G$ the gravitational constant, $c$ is a speed of light and $D$ the distance to a source. The reduced mass quadruple momentum in Cartesian coordinates $\mathbf{x} \equiv [x, y, z]$ is defined as~\cite{Misner:1973prb}:
\begin{equation}
    I_{ij}(t,\mathbf{x})= \int d^3x \rho \left[x_i x_j -\frac{1}{3} \delta_{ij} (x_1^2 + x_2^2 + x_3^2) \right],
\end{equation}
where $\rho$ is the rest mass density.

We consider a rigid body with quadrupole mass moments $I_{ij}$ rotating around the z-axis (no precession or nutation) with a rotational frequency $f_{rot}$. The resulting GW emission can be written, without loss of generality, as~\cite{PhysRevD.20.351, PhysRevD.57.2101, PhysRevD.58.063001}:
\begin{eqnarray}
h_+ &=& \frac{1}{2} h_0 (1 + \cos ^2 \iota) \cos (2 \pi f_0 t), \label{eq:hp}\\
h_\times &=& h_0 \cos \iota \sin(2 \pi f_0 t) \label{eq:hc},
\end{eqnarray}
where $f_0=2f_{rot}$ and the amplitude is:
\begin{equation}
    h_0 = \frac{2}{D} \frac{G}{c^4} \frac{I_{xx}-I_{yy}}{2} (2 \pi f_0)^2.
\end{equation}
Taking an example of a triaxial ellipsoid rotating about a principal axis, the $h_0$ can be expressed as
\begin{equation}
    h_0 = \frac{2}{D} \frac{G}{c^4} \frac{I_{zz} \epsilon}{2} (2 \pi f_0)^2,
    \label{signal_0}
\end{equation}
where the ellipticity is defined as~\cite{PhysRevD.100.024004}:
\begin{equation}
    \epsilon \equiv \frac{I_{xx}-I_{yy}}{I_{zz}}.
    \label{ellipticity}
\end{equation}
The quantity $\epsilon$ is a measure of the core quadrupolar deformation with respect to sphericity. Because GWs considered in this search have a limited duration, we apply a Gaussian envelope to Eqns.~\eqref{eq:hp} and~\eqref{eq:hc} arriving then at Eqns.~\eqref{eqn:sg1} and~\eqref{eqn:sg2}.

This model is a generalization of the one used in previously targeted searches~\citep{SNSearchO1O2,SNSearchS5A5S6}. In that case, the PNS deformation was modeled by a cylinder of mass $M$, length $L$, and radius $R$ rotating about the axis perpendicular to its length. In the current work, no assumption is made about the shape of the deformed star (only that it is rigidly rotating). If one wishes to relate the results for the new model to previous results, it is possible to relate the parameters of the new model ($\epsilon$ and $I_{zz}$) with the ones of the old one ($M$, $L$ and $R$) particularizing the rigid body to a cylinder (note that $f_0$ and $\tau$ are common to both models) such that
\begin{equation}
    \epsilon = \frac{L^2 - 3 R^2}{L^2 + 3 R^2}\qquad,\qquad I_{zz}=\frac{1}{12} M (3R^2+L^2).
\end{equation}
In this search, we use $f_0$ = \{55, 82, 122, 182, 272, 405, 604, 900, 1342, 2000\}\,Hz and $\tau$ = \{0.001, 0.01, 0.1, 1.0\}\,s. The $h_0$ is calculated using $I_{zz}=5.47 \times 10^{45} \mathrm{g\,cm}^2$ that is consistent with the lb4-lb6 waveforms from the previous searches~\cite{SNSearchO1O2,SNSearchS5A5S6}.

The \textit{Torus Fragmentation Instability} scenario \cite{Kobayashi:03,piro:07} proposes that if a black hole and an accretion disk are formed during the collapse, the disk could fragment and large self-gravitating clumps of matter falling into the black hole would produce large amplitude GWs under the appropriate conditions. Fragmentation has been observed in some simplified numerical setups  \cite{Duez:04,giacomazzo:11b} but is very short-lived. However, it is currently unclear if it will develop under more realistic conditions. To model this signal, we employ a simplified model~\cite{santamaria:11dcc} that depends on the mass of the central black hole (BH) $M_\mathrm{BH}=\{5,10\}\,M_\odot$ and the properties of the disk, namely the thickness of the torus $\eta=\{0.3,0.6\}$ and the alpha viscosity parameter $\alpha=0.1$. The torus thickness is defined as $\eta=H/r$, where $H$ is the disk scale height, and $r$ is the local radius. For the disk model considered in~\cite{piro:07}, the mass of the fragmented clump is $M_f=0.53 \eta^3 M_\mathrm{BH}$. The GW amplitude is proportional to the reduced mass of the BH-clump system, $\mu=M_\mathrm{BH} M_f / (M_\mathrm{BH} + M_f)$, which for the parameter space considered here ($M_f \ll M_\mathrm{BH}$) is $\mu \approx M_f$. Following~\cite{SNSearchO1O2}, the piro1 and piro2 stand for a source with $M_\mathrm{BH}=5\,M_\odot$ and $\eta=\{0.3,0.6\}$, respectively. The piro3 and piro4 are for CCSNe with $10\,M_\odot$ black holes and $\eta=\{0.3,0.6\}$, respectively.

\subsubsection{\textit{Ad-hoc} signals}
\label{sec:adhoc}

To constrain the GW energy, luminosity, and PNS ellipticity emitted by a CCSN at a given frequency bin, we use \textit{ad-hoc} sine-Gaussian signals. Currently, the best GW energy constraints are of an order of $10^{-4}\,M_\odot$~\cite{SNSearchO1O2}, and they correspond to the energies of extremely rapidly rotating explosion models. In this search, we use elliptically polarized sine-Gaussians that represent rotating sources. The waveforms are calculated according to the Eqns.~\eqref{eqn:sg1} and~\eqref{eqn:sg2}, and they are parametrized with central frequency $f_0$ = \{55, 82, 122, 182, 272, 405, 604, 900, 1342, 2000\}\,Hz, and $\tau$ = \{0.001, 0.01, 0.1, 1.0\}\,s. The amplitudes do not have physical meaning.

%%%%%%%%%%%%%%%%%%%%%%%%%%%%%%%%%%%%%%%%%%%%%%%%%%%%%
\section{Search Results}
\label{sec:results}

\begin{table}[htb]
\centering
\caption{List of the loudest events for each CCSN.
False alarm rate (FAR) and False Alarm Probability (FAP) for each of them, except SN~2020fqv, which is further analyzed, indicate that they are consistent with background noise.
}
\begin{tabular}{lc@{\hspace*{1em}}c@{\hspace*{1em}}c@{\hspace*{1em}}l}
	\hline
	\hline
	\multicolumn{1}{l}{Supernova} & \multicolumn{1}{c}{Class} & \multicolumn{1}{l}{\,$\eta_{\rm c}$} & \multicolumn{1}{c}{{FAR [Hz]}} & \multicolumn{1}{c}{{FAP}} \\
	\hline
	\hline
	  SN 2019ehk  & \textit{C2} & 5.9 & 1.4e-5 & 0.39\phantom{0} (0.86$\sigma$) \\
	  SN 2019ejj  & \textit{C2} & 6.7 & 1.1e-5 & 0.45\phantom{0} (0.76$\sigma$) \\
	  SN 2019fcn  & \textit{C2} & 6.7 & 1.4e-5 & 0.95\phantom{0} (0.06$\sigma$) \\
	  SN 2019hsw  & \textit{C1} & 5.6 & 4.5e-6 & 0.86\phantom{0} (0.17$\sigma$) \\
	  SN 2020oi   & \textit{C1} & 5.8 & 2.0e-6 & 0.35\phantom{0} (0.93$\sigma$) \\
	  SN 2020cxd  & \textit{C1} & 6.7 & 3.3e-6 & 0.73\phantom{0} (0.34$\sigma$) \\
	  SN 2020dpw  & \textit{C2} & 6.2 & 6.3e-6 & 0.81\phantom{0} (0.23$\sigma$) \\
	  SN 2020fqv  & \textit{C1} & 7.6 & 1.5e-8 & 0.005 (2.78$\sigma$) \\
	\hline
	\hline
\end{tabular}
\label{tab:loudest}
\end{table}

Table~\ref{tab:loudest} presents the search results. The most significant GW candidate is the loudest event of SN~2020fqv with a FAP of 0.54\% ($2.8\sigma$). This event is analyzed in Section~\ref{sec:SN2020fqv} and likely has a noise origin. All other loudest events are consistent with the background.

\subsection{SN~2020fqv loudest event}
\label{sec:SN2020fqv}

\begin{figure}[hbt]
\centering
\includegraphics[width=1.00\columnwidth]{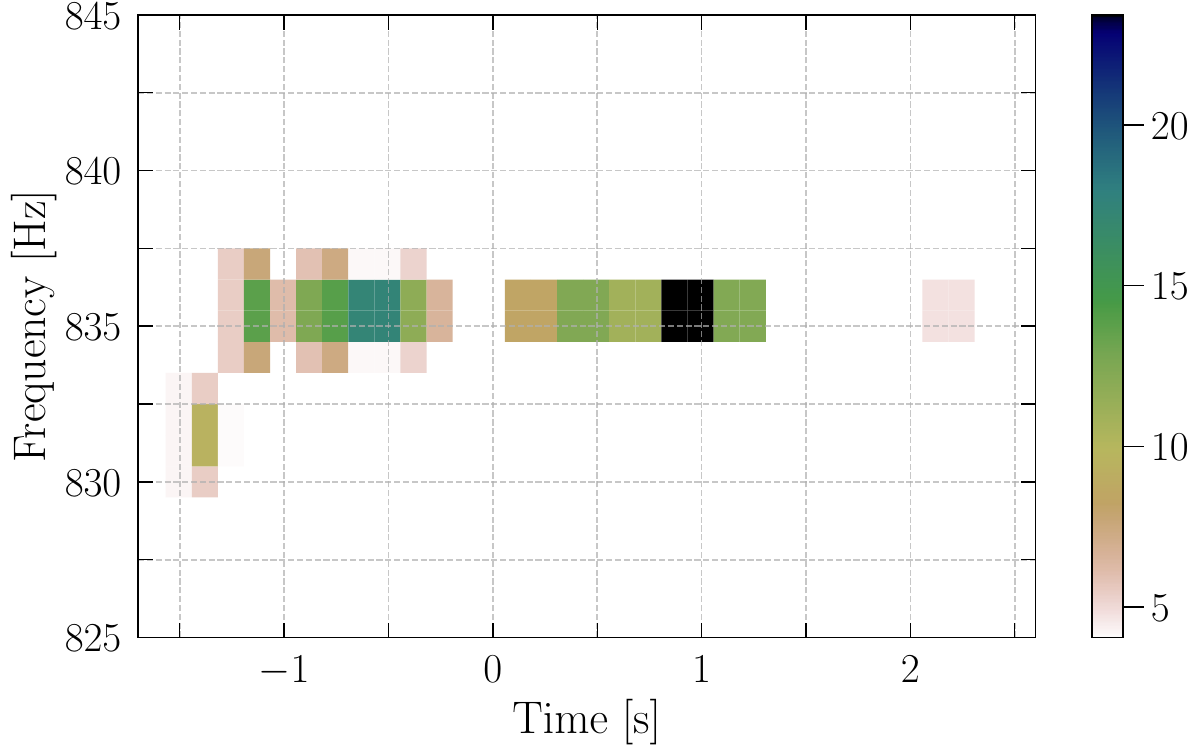}
\caption{
SN~2020fqv loudest event with a $2.8\sigma$ detection significance. The data quality investigations show that this event is most likely of an instrumental origin. The
pixel magnitudes are squared network signal-to-noise ratio.
}
\label{fig:loudestevent}
\end{figure}

Figure~\ref{fig:loudestevent} shows the time-frequency map of the SN~2020fqv loudest event. This is a long (around 4\,s) and narrowband (around 4\,Hz) signal with a peak frequency of 835\,Hz detected with a FAR of 1 per 2.05 years. Given the OSW, the FAP is 0.54\% which corresponds to $2.8\sigma$ confidence. 

Investigating the data quality surrounding this event shows an 837\,Hz noise feature at the LIGO Hanford observatory, ruling this event to be of instrumental origin. The same source of noise has been observed in the O3 search for long-duration transient GW to be instrumental~\cite{KAGRA:2021bhs}.

Apart from the data quality analysis, we consider the search sensitivity to the signals similar to this loudest event. Because it is a narrowband signal, the sine-Gaussians can be used to estimate detectability and the bar model waveforms can be used as probes. Among the analyzed waveforms, the closest is the one at a frequency of 900\,Hz and $\tau= 1$\,s. At the distance of SN~2020fqv (17.3\,Mpc), the detection efficiency is about 23\% (see the next section). Therefore, such an extreme emission can potentially explain the observed SN~2020fqv loudest event. Alternatively, high-frequency modes are also visible in more realistic explosions. For example, a weak feature at 1000\,Hz lasts for 0.5\,s in Figure~6 of Ref.~\citep{Powell:2020cpg}. These signals are at a constant frequency if the mass accretion is low and the PNS isn't changing size.

\begin{figure*}[hbt] 
  \begin{minipage}[c][][t]{0.495\textwidth}
    \vspace*{\fill}
    \flushleft
    \includegraphics[width=0.96\linewidth]{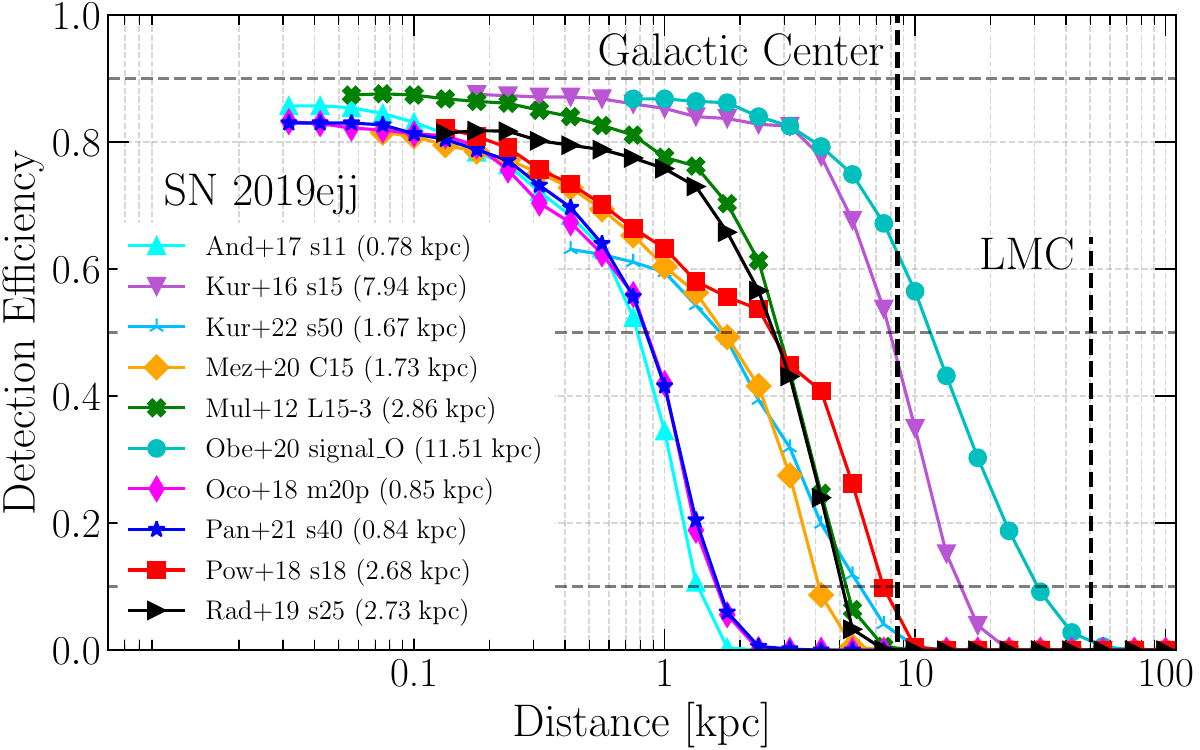}
  \end{minipage}%
  \begin{minipage}[c][][t]{0.495\textwidth}
    \vspace*{\fill}
    \flushright
    \includegraphics[width=0.96\linewidth]{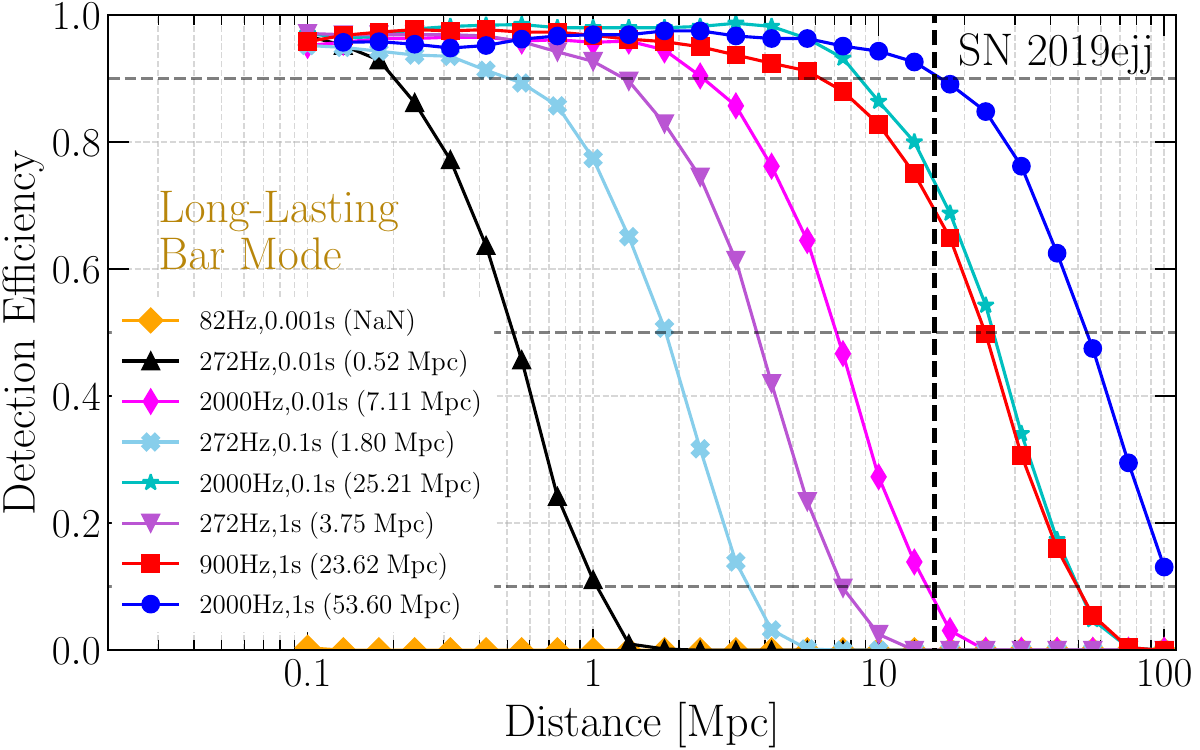}
  \end{minipage}
  \caption{%\label{fig:sn2019ejj}
    The detection efficiency as a function of distance for SN~2019ejj. The numbers in the brackets are distances at 50\% detection efficiencies. Horizontal dashed lines show 10\%, 50\%, and 90\% detection efficiencies. \textit{Left panel} shows the efficiencies for 12 CCSN models derived from multidimensional CCSN simulations. As references, the Galactic Center and Large Magellanic Cloud (LMC) distances are plotted. \textit{Right panel} provides the detection efficiencies for the extreme emission long-lasting bar mode model. Some models are reaching the distance of SN~2019ejj. Given a null detection, it allows excluding parameter spaces of this extreme emission model as discussed in Sec.~\ref{sec:constraints}.
     }
  \label{fig:effcurves}
\end{figure*}

\begin{table*}[htb]
\centering
\caption{Distance (in kpc) of the 50\% detection efficiency reached for each CCSN with neutrino-driven explosions, magnetorotational explosions (signal\_O), black hole formation (s40), and phase transitions (s50) models. Values in bold represent the farthest distance reach for each model. Mark `-' means that the detection efficiency did not reach 50\%.\\}

%\begin{tabular}{lc@{\hspace*{1em}}c@{\hspace*{1em}}c@{\hspace*{1em}}l}
\begin{tabular}{ lc@{\hspace*{1em}}c@{\hspace*{1em}}c@{\hspace*{1em}}c@{\hspace*{1em}}c@{\hspace*{1em}}c@{\hspace*{1em}}c@{\hspace*{1em}}c@{\hspace*{1em}}c@{\hspace*{1em}}c@{\hspace*{1em}}c@{\hspace*{1em}}c@{\hspace*{1em}}c@{\hspace*{1em}}c@{\hspace*{1em}}c@{\hspace*{1em}} }
	\hline
	\hline
	\multicolumn{1}{l}{Supernova} & 
	\multicolumn{1}{c}{And+17} & 
	\multicolumn{1}{c}{Kur+16} & 
	\multicolumn{1}{c}{Kur+22} & 
	\multicolumn{1}{c}{Mez+20} & 
	\multicolumn{1}{c}{Mul+12} & 
	\multicolumn{1}{c}{Obe+20} & 
	\multicolumn{2}{c}{Oco+18} & 
	\multicolumn{1}{c}{Pan+21} & 
	\multicolumn{2}{c}{Pow+18} & 
	\multicolumn{3}{c}{Rad+19} \\
	\multicolumn{1}{l}{} & \multicolumn{1}{c}{s11} & \multicolumn{1}{c}{s15} & \multicolumn{1}{c}{s50} & \multicolumn{1}{c}{C15} & \multicolumn{1}{c}{L15-3} & \multicolumn{1}{c}{signal\_O} & \multicolumn{1}{c}{m20} & \multicolumn{1}{c}{m20p} & \multicolumn{1}{c}{s40} & \multicolumn{1}{c}{s18} & \multicolumn{1}{c}{s3.5} & \multicolumn{1}{c}{s9} & \multicolumn{1}{c}{s13} & \multicolumn{1}{c}{s25} \\
	\hline
	\hline
	  SN 2019ehk  &\,\,-             &6.57           &-             &0.52          &2.47          &4.22           &0.18          &0.77          &0.38          &3.05          &1.54          &\,\,0.16          &0.33          &3.11         \\
	  SN 2019ejj  &\,\,0.78          &7.94           &1.67         &1.73          &2.86 &11.51          &0.64          &0.85          &0.84          &2.68          &1.79          &\,\,0.26 &0.61 &2.73         \\
        SN 2019fcn  &\,\,0.58          &7.40           &0.80          &0.84          &2.46          &8.81           &0.50          &0.64          &0.58          &0.83          &0.87          &\,\,0.22          &0.49          &1.86         \\
        SN 2019hsw  &\,\,0.70          &5.60           &1.82          &2.24          &2.33          &13.40          &0.60          &0.76          &0.77          &3.85          &2.04          &\,\,0.17          &0.49          &2.82         \\
	  SN 2020oi   &\,\,0.63          &6.53           &-             &1.15          &2.36          &9.52           &0.56          &0.70          &0.61          &1.71          &0.94          &\,\,0.21          &0.52          &1.96         \\
	  SN 2020cxd  &\,\,\textbf{0.88} &\textbf{8.90}  &\textbf{2.13} &\textbf{2.74} &\textbf{3.17}          &\textbf{14.65} &\textbf{0.74} &\textbf{0.95} &\textbf{0.94} &\textbf{4.74} &\textbf{2.38} &\,\,\textbf{0.27}          &\textbf{0.67}          &\textbf{3.15}\\
	  SN 2020dpw  &\,\,0.79          &8.66           &1.70          &2.46          &2.96          &13.43          &0.68          &0.85          &0.90          &4.30          &2.24          &\,\,0.27          &0.61          &2.86         \\
	  SN 2020fqv  &\,\,0.73          &6.86           &1.56          &2.38          &2.53          &13.42          &0.65          &0.82          &0.81          &4.17          &2.17          &\,\,0.21          &0.55          &2.90         \\
	\hline
	\hline
\end{tabular}
\label{tab:Distances50eff}
\end{table*}

%%%%%%%%%%%%%%%%%%%%%%%%%%%%%%%%%%%%%%%%%%%%%%%%%%%%%
\subsection{Detection efficiency vs distance}
\label{sec:eff}
%%%%%%%%%%%%%%%%%%%%%%%%%%%%%%%%%%%%%%%%%%%%%%%%%%%%%

To assess the search sensitivity, we produce detection efficiency vs distance for a wide range of CCSN models. Figure~\ref{fig:effcurves} shows the detection efficiencies of the models described in \ref{sec:wave} for the sky location and OSW of SN~2017ejj. The numbers in parentheses are detection reaches. As references, the left plot shows the distances to the Galactic Center ($\sim$8.5\,kpc) and the Large Magellanic Cloud (49.6\,kpc~\cite{Pietrzyski2019}) that hosted SN~1987A. Table~\ref{tab:Distances50eff} summarized the distance reached for all analyzed CCSNe. For each model, the values are consistent across the CCSNe. The largest distances are mostly achieved with SN~2020cxd.

The neutrino-driven explosions (models s11, s15, C15, L15-3, m20, m20p, s18, s3.5, s9, s13, and s25) are detectable up to a few kpc, with the s15 model being reached to the Galactic Center. When comparing the three models with solar metallicity progenitor stars (s9, s13, and s25), the distance reach increases with progenitor mass.

The magnetorotationally-driven explosion model signal\_O has a larger distance reach compared to the neutrino-driven explosions. For most CCSNe, the distances exceed the Galactic Center; for a few of them, the detection probability is non-zero at the distance of the Large Magellanic Cloud. The quantum chromodynamics phase transition explosions (s50) can be reached up to around 2.1\,kpc

The right panel of Figure~\ref{fig:effcurves} shows SN~2017ejj detection efficiencies of a few GW signals from the extreme emission bar model. The detection reaches increase with signals' duration and peak frequency, up to tens of Mpc. For example, 82\,Hz and 1\,ms signals are not detectable at 0.1\,Mpc. On the contrary, 900\,Hz and 1\,s signals can reach distances of the analyzed supernovae. In particular, the relatively high detection efficiency is achieved for the SN~2020fqv.

\begin{figure*}[hbt] 
  \begin{minipage}[c][][t]{0.495\textwidth}
    \vspace*{\fill}
    \flushleft
    \includegraphics[width=0.96\linewidth]{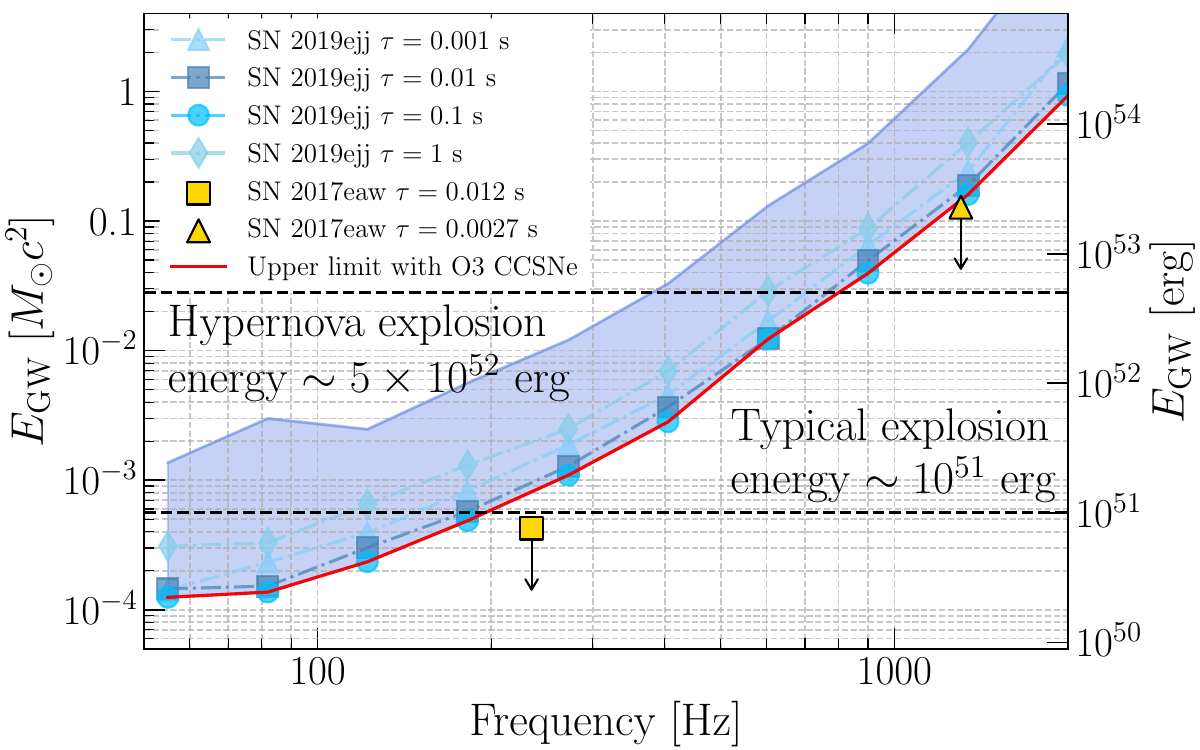}
  \end{minipage}%
  \begin{minipage}[c][][t]{0.495\textwidth}
    \vspace*{\fill}
    \flushright
    \includegraphics[width=0.96\linewidth]{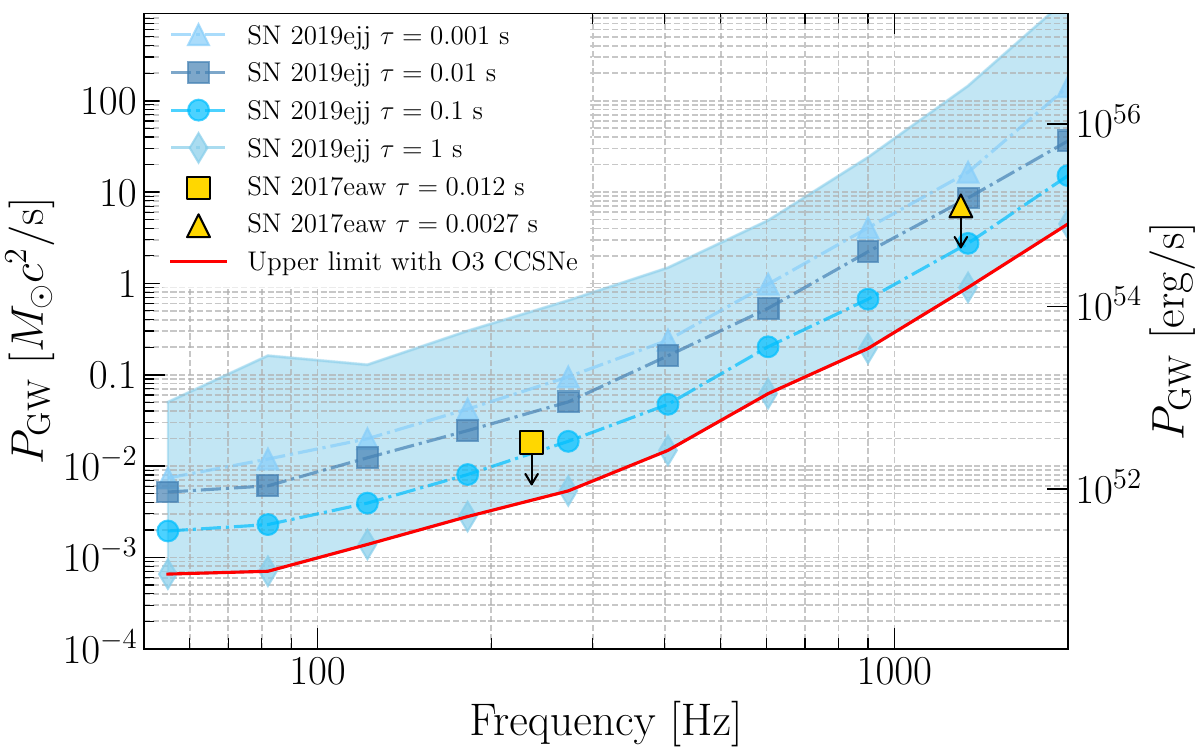}
  \end{minipage}
  \caption{\label{fig:energy}
     The upper limits on the GW energy ($E_\mathrm{GW}$) and luminosity (or power, $P_\mathrm{GW}$) emitted by a CCSN engine. The shaded region contains combined results from all analyzed CCSNe. The tightest results are obtained for SN~2019ejj. At 50\,Hz the stringent energy constraints are $10^{-4}\,M_\odot c^2$ for signals $1-100$\,ms. The best upper limits for GW luminosity are $5 \times 10^{-4}\,M_\odot c^2 /\mathrm{s}$ for signals at 50\,Hz and 1\,s long. Our results are around two times less stringent than those obtained with SN~2017eaw~\cite{SNSearchO1O2}. The $E_\mathrm{GW}$ upper limits are still much higher than those derived from numerical simulations~\cite{Szczepanczyk:2021bka}.
     }
  \label{fig:eff}
\end{figure*} 

We use the L15-3 model to compare the search sensitivity between the previous searches~\cite{SNSearchO1O2,SNSearchS5A5S6} and this search. In Ref.~\cite{SNSearchO1O2} a typical distance was around 1.5\,kpc with a maximum of 2.4\,kpc. Here, the distances are above 2\,kpc with a maximum of 3.44\,kpc. This corresponds to an improvement in the detectors' sensitivities of around 50\%.

Detection distances were recently derived in the O3 all-sky generic LIGO-Virgo-KAGRA search~\cite{KAGRA:2021tnv}. They share a common model with this work: the m20 simulation. They estimate the distance at a 10\% detection efficiency to be $\sim$1\,kpc. It is calculated at FAR of 1/100~years. In this search, the distance at a 10\% detection efficiency is estimated at around 2\,kpc but with a higher FAR associated with the loudest event and using a shorter time window. As explained in Sec.~\ref{sec:cwb}, better sensitivities are achieved with targeted searches.

%%%%%%%%%%%%%%%%%%%%%%%%%%%%%%%%%%%%%%%%%%%%%%%%%%%%%
\section{CCSN engine constraints}
\label{sec:constraints}

Because no GW is found, we constrain the dynamics of the CCSN engine, similarly to Ref~\cite{SNSearchO1O2}. However, we significantly expand the statements. Rather than probing low- and high-frequency GW emission, a broad frequency spectrum is explored here. Also, the analyzed range of signal durations expanded. The constraints presented in this section can be divided into generic and modeled. The statements on the generic constraints include the upper limits on emitted GW energy, luminosity, and PNS ellipticity. Following~\cite{SNSearchO1O2}, we further constrain the parameter spaces of the extreme emission models. However, the bar model is generalized compared to~\cite{SNSearchO1O2}.

%%%%%%%%%%%%%%%%%%%%%%%%%%%%%%%%%%%%%%%%%%%%%%%%%%%%%%%%%%%%%%%%%%
\subsection{Upper limits on GW energy and luminosity}
\label{sec:egw}
%%%%%%%%%%%%%%%%%%%%%%%%%%%%%%%%%%%%%%%%%%%%%%%%%%%%%%%%%%%%%%%%%%%

Similarly to~\cite{SNSearchO1O2}, we constrain the GW energy ($E_\mathrm{GW}$) emitted by a CCSN engine. Additionally, we explore CCSN explosion's dynamics with GW luminosity (or power, $P_\mathrm{GW}$). Assuming a rotating CCSN source, the total energy emitted in GWs is~\cite{Sutton:2013ooa}:
\begin{equation}
  E_\mathrm{GW} = \frac{2}{5} \frac{\pi^2 c^3}{G}D^2 f^2_0 h_\mathrm{rss50}^2 \, ,
  \label{eqn:energy}
\end{equation}
where $f_0$ is the peak frequency, $D$ is the distance to the source and $h_\mathrm{rss50}$ is an $\hrss$ (Eqn.~\eqref{eqn:hrss}) at 50\% detection efficiency. The detection efficiencies versus $\hrss$ are produced with elliptically polarized sine-Gaussians described in Sec.~\ref{sec:adhoc}.

The GW luminosity is the ratio between emitted GW energy and the duration of this emission. The detector Gaussian noise and glitches affect the events' reconstructed parameters, such as duration. To minimize this bias, we use signal duration containing 90\% of the signal's energy. In terms of the $\tau$ parametrization, the cWB reconstructed duration can be approximated by $1.65 \tau_\mathrm{rec}$. The GW power is then defined as:
\begin{equation}
  P_\mathrm{GW} = \frac{0.9 E_\mathrm{GW}}{1.65 \tau_\mathrm{rec}}\,.
\end{equation}

The left panel of Figure~\ref{fig:energy} shows the $E_\mathrm{GW}$ constraints. The shaded region contains combined results from all analyzed CCSNe. From the optical observations, the typical CCSN explosion energy is around $10^{51}$\,erg, while for hypernovae, it can be even $5 \times 10^{52}$\,erg~\cite{NOMOTO2010191,Tanaka:2008ss,2011A&A...532A.100U}. The tightest results are obtained primarily for SN~2019ejj. At 50\,Hz the stringent energy constraints are $10^{-4}\,M_\odot c^2$ for signals $1-100$\,ms. The constraints obtained in the previous search~\cite{SNSearchO1O2} are also shown in the figure. At 235\,Hz the GW emission was estimated with $\tau=12$\,ms, while at 1034\,Hz it was $\tau=2.7$\,ms. The constraints with SN~2019ejj are around two times less stringent, mainly due to a larger distance.

The right panel of Figure~\ref{fig:energy} reports on the emitted GW power. The tightest results are obtained for SN~2019ejj. The shaded region contains combined results from all analyzed CCSNe. The stringent power constraints are $6 \times 10^{-4}\,M_\odot c^2$/s for signals at 50\,Hz and 1\,s long. We re-analyzed SN~2017eaw~\cite{SNSearchO1O2}, and we find that the constraints with SN~2019ejj are around a factor of two less stringent.

%%%%%%%%%%%%%%%%%%%%%%%%%%%%%%%%%%%%%%%%%%%%%
\subsection{Upper limits on PNS ellipticity}
\label{sec:ell}
%%%%%%%%%%%%%%%%%%%%%%%%%%%%%%%%%%%%%%%%%%%%%

\begin{figure}[hbt]
  \centering
 \includegraphics[width=1.00\columnwidth]{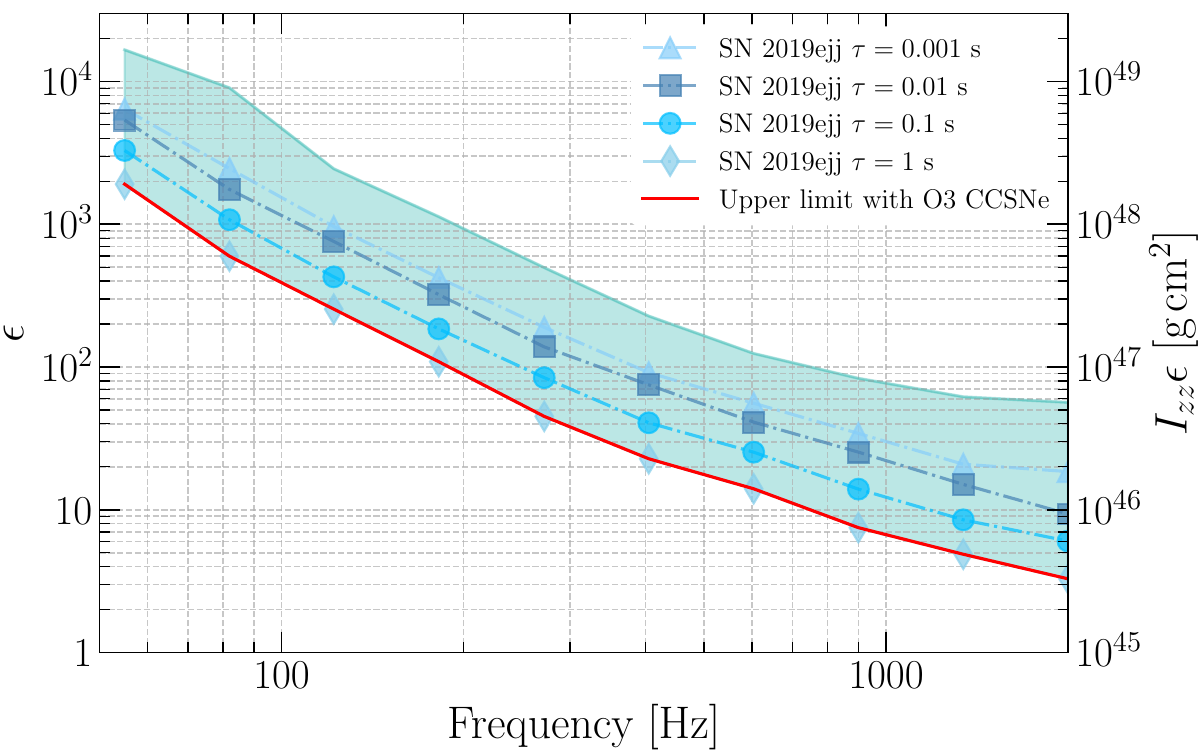}
  \caption{
  The upper limits on the PNS ellipticity. Assuming a principal canonical moment of inertia for neutron stars, $I_{zz}=10^{45} \mathrm{g\,cm}^2$, the stringent upper limits on the ellipticities are down to around 3 at 2\,kHz.
  }
\label{fig:ell}
\end{figure}

The bar model described in Sec.~\ref{sec:extreme} can be used to provide upper limits on the allowed ellipticities of the core deformations. The angle-averaging of $h_{rss}$ gives $h_{rss}^2 = h_0^2 \sqrt{\pi/2} \tau$ and one can arrive to an experimental expression of $I_{zz}\epsilon$:
\begin{equation}
    I_{zz} \epsilon = \frac{Dc^4}{G (2 \pi f_0)^2} \left(\frac{2}{\pi \tau^2_\mathrm{rec}}\right)^{1/4} h_{\mathrm rss50}\,,
    \label{eqn:ell}
\end{equation}
where $h_\mathrm{rss50}$ is an $h_\mathrm{rss}$ value at 50\% detection efficiency, and $\tau_\mathrm{rec}$ is signal duration, both are estimated from cWB. Figure~\ref{fig:ell} reports upper limits on the ellipticity for a range of GW signal frequencies and durations. The shaded region contains combined results from all analyzed CCSNe. The degree of deformation $\epsilon$ can also be presented assuming a principal canonical moment of inertia for neutron stars, $I_{zz}=10^{45} \mathrm{g\,cm}^2$~\cite{Pisarski:2019vxw}. The stringent upper limits on ellipticity are obtained for the signals with $\tau=1$\,s, ranging from $10^4$ at the lowest search frequency to 3.3 at 2\,kHz. The $\epsilon$ values increase with shorter signals. 
% If bars are created in CCSNe, they are rather short-lived.

%%%%%%%%%%%%%%%%%%%%%%%%%%%%%%%%%%%%%%%%%%%%%
\subsection{Model exclusion statements}
\label{sec:mep}
%%%%%%%%%%%%%%%%%%%%%%%%%%%%%%%%%%%%%%%%%%%%%

The previous search~\cite{SNSearchO1O2} excluded the parameter spaces of two extreme emission models. Here, we continue this effort. The model exclusion probability of combined $N$ CCSNe is calculated as~\cite{kalmus:13}:
\be
P_\mathrm{excl} = 1 - \prod_{i =1}^N (1-\varepsilon_i(D_i))\,. \label{eq:reach}
\ee  
The $\varepsilon(D) = a \times \mathcal{E}(D)$ is a detection efficiency $\mathcal{E}(D)$ reduced by the coverage duty factor $a=T_\mathrm{coinc} / \Delta t$ (see Table~\ref{tab:sne}). 

\begin{figure}[hbt]
  \centering
  \includegraphics[width=1.00\columnwidth]{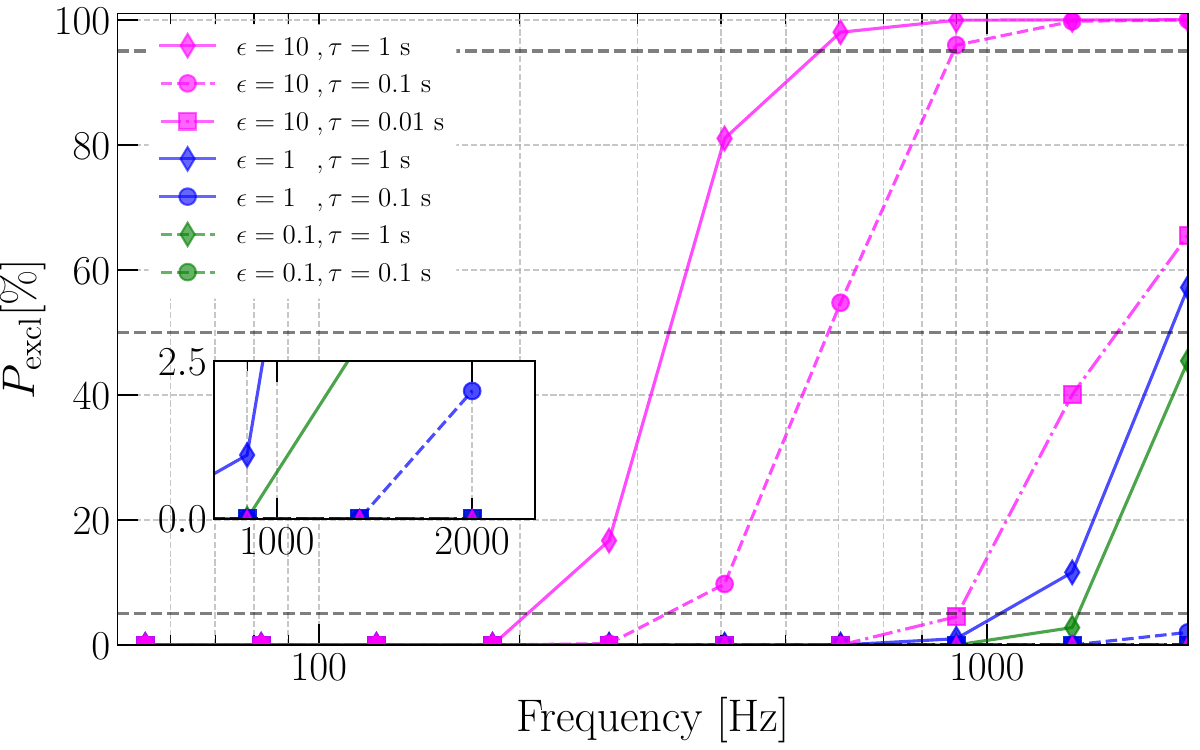}
  \caption{
  Model exclusion probability $P_\mathrm{excl}$ for long-lasting bar mode instability model. The numbers are calculated by accumulating results from CCSNe in O1, O2, and O3. The GW emissions from bars with $\epsilon=10$ are excluded at almost 100\% confidence above 900\,Hz for $\tau=1$\,s and $\tau=0.1$\,s. The probabilities decrease with signal ellipticities and durations. The emissions with the ellipticity of 0.1 and $\tau=1$\,s are excluded up to around 50\%. GW emission with $\tau=1$\,ms cannot yet be excluded.
  }
\label{fig:mep}
\end{figure}

\begin{figure*}[hbt] 
  \begin{minipage}[c][][t]{0.495\textwidth}
    \vspace*{\fill}
    \flushleft
    \includegraphics[width=0.96\linewidth]{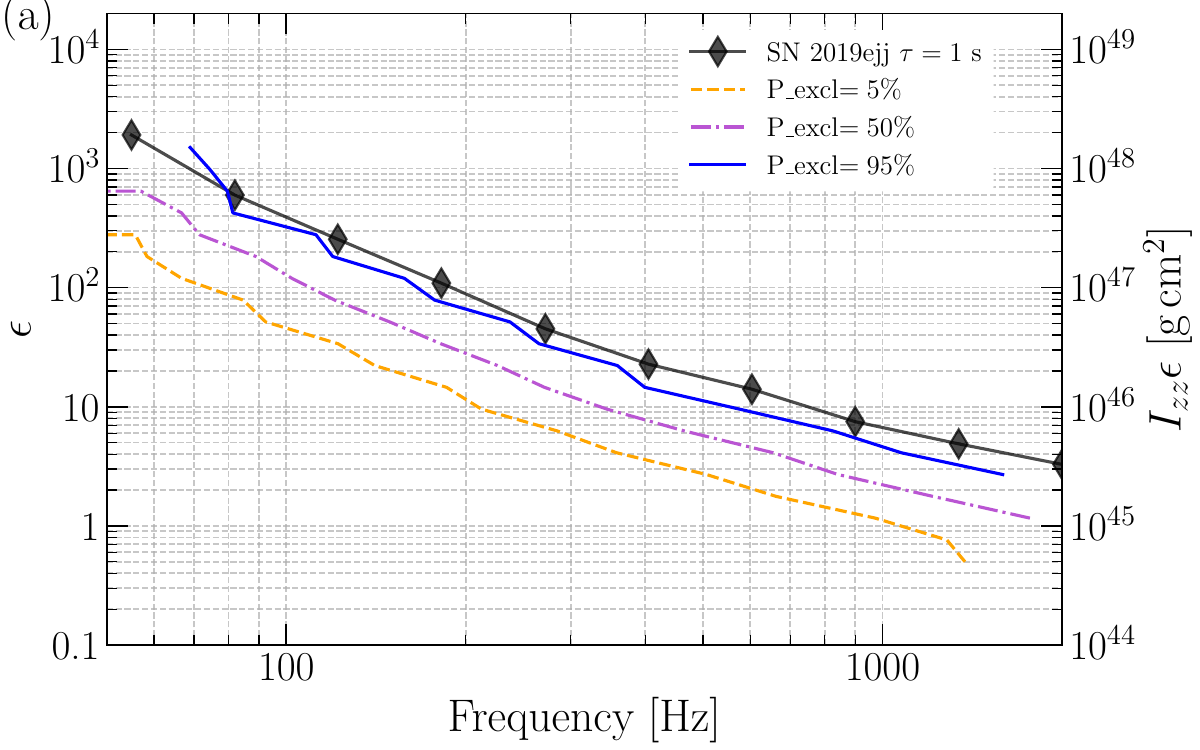}
    \includegraphics[width=0.96\linewidth]{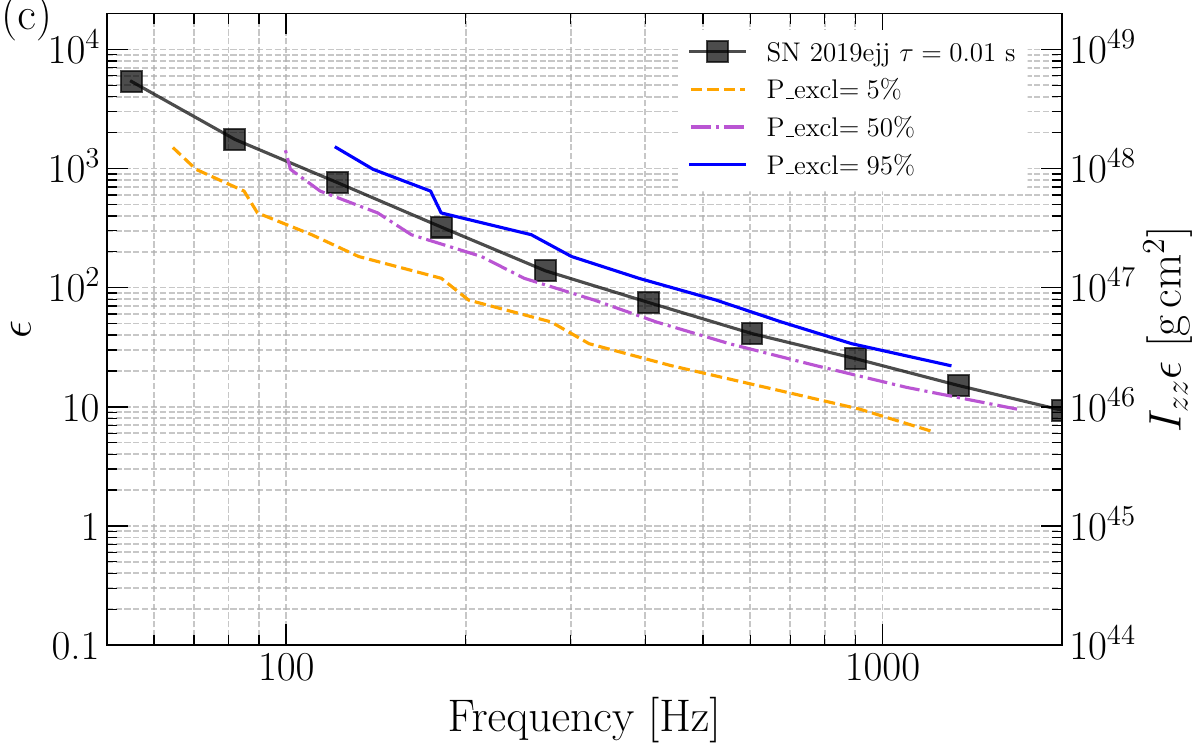}
  \end{minipage}%
  \begin{minipage}[c][][t]{0.495\textwidth}
    \vspace*{\fill}
    \flushright
    \includegraphics[width=0.96\linewidth]{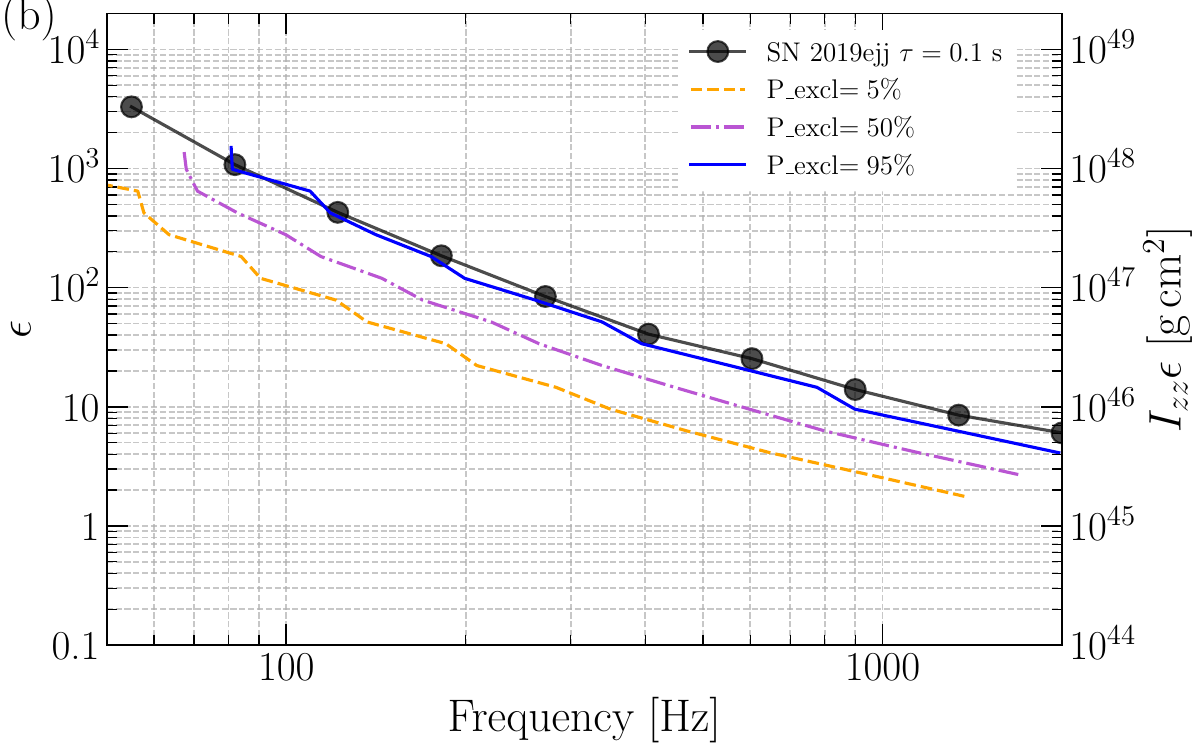}
    \includegraphics[width=0.96\linewidth]{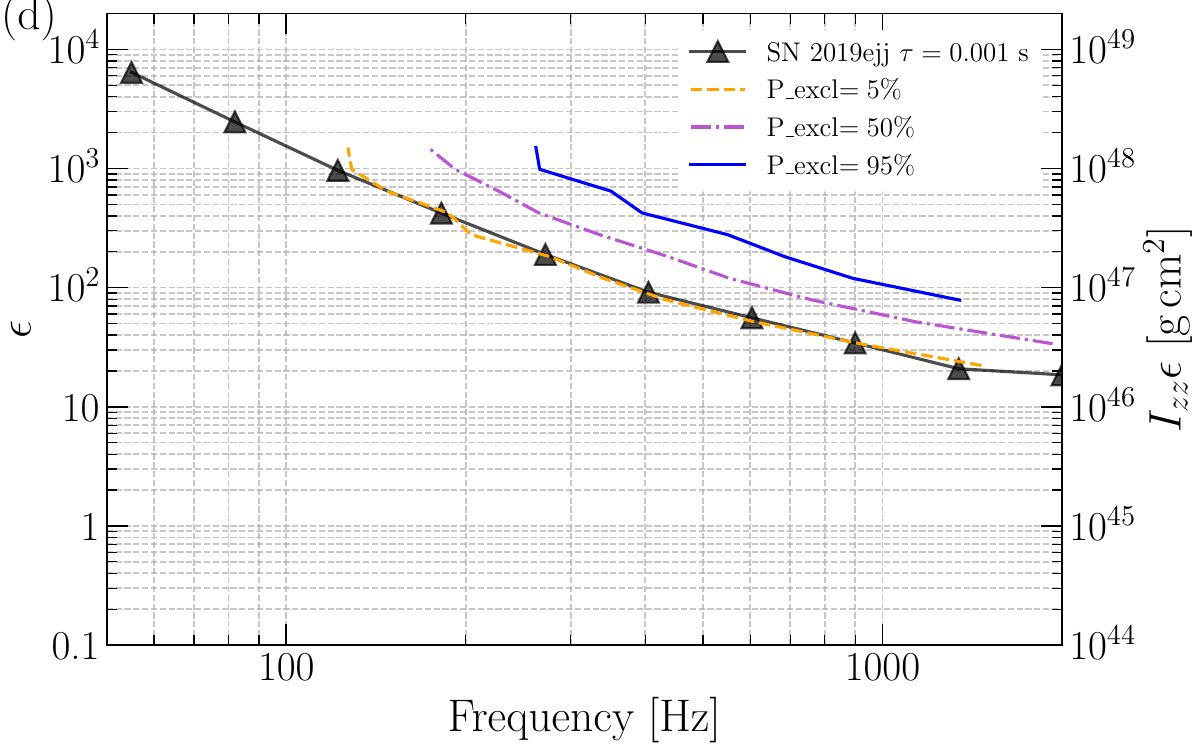}
  \end{minipage}

  \caption{
     The constraint of the PNS ellipticity for the long-lasting bar model using a population of the CCSNe analyzed in this and the previous~\cite{SNSearchO1O2} search. The best constraints are obtained for long signals and GW emission at 2\,kHz. These observational ellipticity constraints are around an order of magnitude less stringent than the ellipticities of the CCSN simulations~\cite{Bugli:22,Shibagaki:20}.
     }
  \label{fig:ell_pop}
\end{figure*}

The previous search~\cite{SNSearchO1O2} assumed a standard candle approach (the CCSN source is optimally oriented). Here, this approach is generalized to all possible source orientations. The CCSNe from O1 and O2 are then re-analyzed, and Figure~\ref{fig:mep} shows the model exclusion probabilities combining results from O1, O2, and O3 CCSNe. The $P_\mathrm{excl}$ values are comparable between combined O1-O2 results~\cite{SNSearchO1O2} and this search. The model exclusion probabilities decrease with the PNS ellipticities. For $\epsilon=10$, the emissions of $\tau=1$\,s and $\tau=0.1$\,s are excluded up to 100\% for high-frequency emission (900\,Hz and above). For emissions with $\tau=0.1$\,s at 2\,kHz are excluded at around 50\%. The emissions with $\tau=0.1$\,s and ellipticities of 1 and 0.1 cannot be yet reliably excluded, but they are non-zero. Finally, for $\tau=0.001$\,s, the emissions cannot be yet constrained. Interpreting these results, if bars are created in CCSNe, they are rather short-lived.

The analysis with O3 CCSNe allows further constraining of the fragmentation instability model (Sec.~\ref{sec:extreme}) analyzed in~\cite{SNSearchO1O2}. The model exclusion probabilities for piro2 and piro4 obtained with this search are 10.3\% and 93.8\%. The cumulative $P_\mathrm{excl}$ values with O1, O2, O3 CCSNe for piro1-piro4 are 0, 41.2\%, 5.2\% and 99\%. These results further assure us that if clumps of matter are formed in type-II and type-Ib/c supernovae, these clumps are small compared to the central black holes. Additionally, if the torii are created around black holes, they are either non-fragmented or rather thin.

%%%%%%%%%%%%%%%%%%%%%%%%%%%%%%%%%%%%%%%%%%%%%
\subsection{Ellipticity constraints for the bar model}
\label{sec:ell_pop}
%%%%%%%%%%%%%%%%%%%%%%%%%%%%%%%%%%%%%%%%%%%%%

Section~\ref{sec:ell} provides generic upper limits of the ellipticity based on single CCSN targets using ad-hoc sine-Gaussians. Here, a population of CCSNe used in this and the previous search~\cite{SNSearchO1O2} is used to find upper limits on the ellipticity for the long-lasting bar model. These constraints are derived from the model exclusion statements. Because the signal amplitudes are proportional to the ellipticity (Eqn.~\ref{signal_0}), we can probe the continuous values of the ellipticities. This method provides constraints at chosen confidence levels; here, we take 5\%, 50\%, and 95\%.

Figure~\ref{fig:ell_pop} shows the ellipticity constraints for the four signal durations. Compared with the generic ellipticity constraints, these upper limits are more stringent for $\tau$ of 1\,s, 0.1\,s, and 0.01\,s, and are less stringent for shorter signals. At 2\,kHz GW emission, we constrain the core deformation to $\epsilon=1$. Note that for a bar model, the frequency of GW emission is twice the rotational frequency. These constraints do not span the full frequency band because some GW signals have very low detection efficiencies at smaller analyzed distances. See Figure~\ref{fig:effcurves} for 82\,Hz and 1\,ms example signal with a zero detection efficiency across the analyzed source distances. In such a case, it's not possible to achieve even $P_\mathrm{excl}=5\%$.

Recent CCSN simulations show bar mode instabilities appearing with an amplitude of $10^{-21}$ at around 300\,Hz~\cite{Bugli:22} or $10^{-20}$ at around 800\,Hz~\cite{Shibagaki:20}. Taking $I_{zz}\epsilon = 0.1\times 10^{45} \mathrm{g\,cm}^2$, or $\epsilon=0.1$ for a canonical moment of inertia for neutron stars, one gets similar amplitudes at these two frequencies using the bar model. It is around an order of magnitude from the obtained upper limits.

%%%%%%%%%%%%%%%%%%%%%%%%%%%%%%%%%%%%%%%%%%%%%%%%%%%%%
\section{Summary and Discussion}
\label{sec:summaries}
%%%%%%%%%%%%%%%%%%%%%%%%%%%%%%%%%%%%%%%%%%%%%%%%%%%%%

We present the results of a search for GWs in coincidence with CCSNe observed optically during the third observing run of the Advanced LIGO and Advanced Virgo detectors. For eight CCSNe, all within a distance of around 30\,Mpc, we calculated windows where a GW transient could be found: SNe 2019ehk, 2019ejj, 2019fcn, 2019hsw, 2020oi, 2020cxd, 2020dpw, 2020fqv. The loudest event of SN~2020fqv has a significance of $2.8\sigma$, but the detailed analysis indicates that this event most likely has a noise origin. The loudest events from all other CCSNe are consistent with the background.

%%%%% -> Distances
For the analyzed waveforms from multidimensional CCSN simulations, the distances at 50\% detection efficiency do not reach beyond Milky Way. For neutrino-driven explosions, these distances are up to 8.9\,kpc. Because the GW emissions are typically larger for magnetorotationally-driven explosions, the distance reaches are further, up to 14.7\,kpc for the analyzed model. For the selected black hole formation and quantum chromodynamics phase transition models they are up to 0.9\,kpc and 2.1\,kpc, respectively. However, the distances for extreme emission models can be further than those of the analyzed CCSNe. This allows further constraining of the CCSN engine.

%%%%% -> Energy and Power
We provide generic CCSN engine constraints and those for the extreme emission models. The generic constraints include upper limits on the GW energy, and, for the first time, GW luminosity and PNS ellipticity. The analysis is performed across a wide frequency range from 50\,Hz to 2\,kHz. At frequencies less than 900\,Hz the obtained energies are below $10^{51}$\,erg (a typical CCSN explosion energy). The upper limits of around $10^{-4}\,M_\odot c^2$ are at frequencies below 100\,Hz for signals with durations $1-100$\,ms. After re-analyzing SN~2017eaw~\cite{SNSearchO1O2}, our constraints are less stringent by a factor of around two.

%%%%% -> Ellipticity
We report generic constraints of the PNS ellipticity, $I_{zz}\epsilon$. Assuming the principal canonical moment of inertia for a neutron star ($I_{zz}=10^{45} \mathrm{g\,cm}^2$), the upper limits for $\tau=1$\,s are down to $\epsilon=3$ for GW signals 2\,kHz and they increase above $10^3$ for the lowest frequencies. These upper limits are becoming less stringent for shorter signal durations.

%%%%% -> Bar model
By combining the results obtained with the data from O1 and O2~\cite{SNSearchO1O2}, we improve the constraints of the parameter spaces of the extreme emission models assuming a standard candle approach. Specifically, the long-lasting bar mode models are analyzed in more detail compared with~\cite{SNSearchO1O2}. The most stringent constraints are at high frequencies, down to $\epsilon=1$ for 1\,s long emission. These constraints are less stringent for shorter signals. We note that the obtained ellipticities are roughly an order of magnitude larger than those obtained from the recent CCSN simulations~\cite{Bugli:22,Shibagaki:20}.

% Summary
The targeted search with O1-O2 data~\cite{SNSearchO1O2} allowed, for the first time, constraining the CCSN engine. While this search has not improved the upper limits on the GW energy emission, the upper limits on GW luminosity and PNS ellipticity are reported for the first time. By combining O1, O2, and O3 data, the extreme emission models are constrained further compared to~\cite{SNSearchO1O2}. The obtained ellipticities of the rotating cores are around an order of magnitude above the largest obtained in CCSN simulations. Future observing runs with improved sensitivities have the potential to accumulate enough statistics to constrain the CCSN simulations. Moreover, our results indicate that near-future data might be able to observationally constrain other CCSN models, such as core fragmentation or higher $T/|W|$ effects, and others.

%%%%%%%%%%%%%%%%%%%%%%%%%%%%%%%%%%%%%%%%%%%%%%%%%%%%%
\begin{acknowledgments}\label{sec:acknowledgments}
%%%%%%%%%%%%%%%%%%%%%%%%%%%%%%%%%%%%%%%%%%%%%%%%%%%%%

This document has been assigned LIGO Laboratory document number P2200361. This research has made use of data, software, and/or web tools obtained from the Gravitational Wave Open Science Center, a service of LIGO Laboratory, the LIGO Scientific Collaboration, and the Virgo Collaboration. This material is based upon work supported by NSF’s LIGO Laboratory which is a major facility fully funded by the National Science Foundation. The work by SK was supported by NSF Grant No. PHY 2110060. MZ was supported by NSF Grant No. PHY-1806885. MC and YZ are partially supported by NSF award PHY-2011334. AS is supported by NSF Grant No. PHY-2110157. JP is supported by the Australian Research Council (ARC) Discovery Early Career Researcher Award (DECRA) project number DE210101050 and the ARC Centre of Excellence for Gravitational Wave Discovery (OzGrav) project number CE170100004. Antelis and Moreno’s research is partially supported by CONACyT Ciencia de Frontera Project No. 376127. This work was partially supported by the: Polish National Science Centre grants No. 2017/26/M/ST9/00978, 2022/45/N/ST9/04115 and 2023/49/B/ST9/02777, POMOST/2012-6/11 Program of Foundation for Polish Science co-financed by the European Union within the European Regional Development Fund. The project is co-financed by the Polish National Agency for Academic Exchange within Polish Returns Programme. PC, JF and MO are supported by the grants PGC2018-095984-B-I00, PID2021-125485NB-C21 and PID2021-127495NB-I00 of the Spanish Agencia Estatal de Investigación and PROMETEO/2019/071 of the Generalitat Valenciana, all funded by the MCIN and the European Union. MO was supported by the Spanish Ramon y Cajal programme (RYC-2018-024938-I). QLN was supported in part by the NSF Grant No. PHY-1748958. The authors would like to thank the DLT40 and ASASSN teams for monitoring the sky for the purpose of this search. The authors would like to thank Noel Richardson for suggesting the Kepler light curves as testing ground for the CCSN light curve interpolations.

%%%%%%%%%%%%%%%%%%%%%%%%%%%%%%%%%%%%%%%
\end{acknowledgments}
%%%%%%%%%%%%%%%%%%%%%%%%%%%%%%%%%%%%%%%%

\appendix

\section{Calibration Error}
\label{sec:cal}

\begin{figure*}[hbt]
  \begin{minipage}[c][][t]{0.495\textwidth}
    \vspace*{\fill}
    \flushleft
    \includegraphics[width=0.96\linewidth]{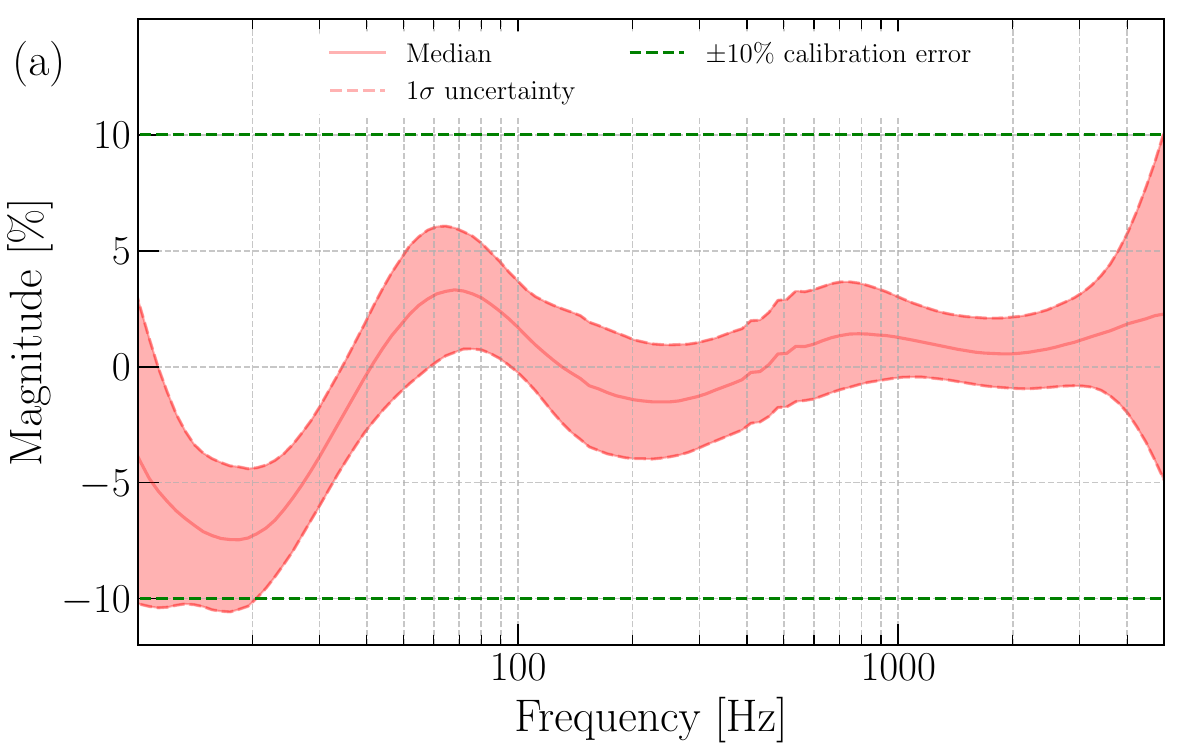}
    \includegraphics[width=0.96\linewidth]{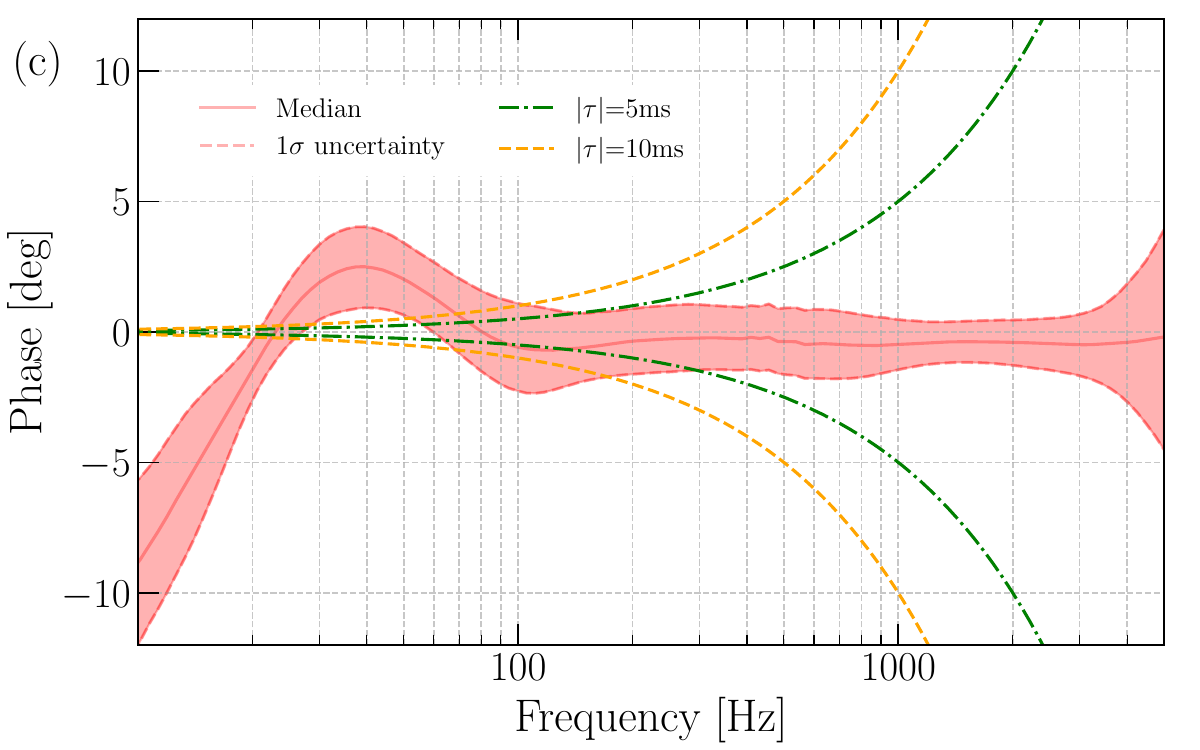}
  \end{minipage}%
  \begin{minipage}[c][][t]{0.495\textwidth}
    \vspace*{\fill}
    \flushright
    \includegraphics[width=0.96\linewidth]{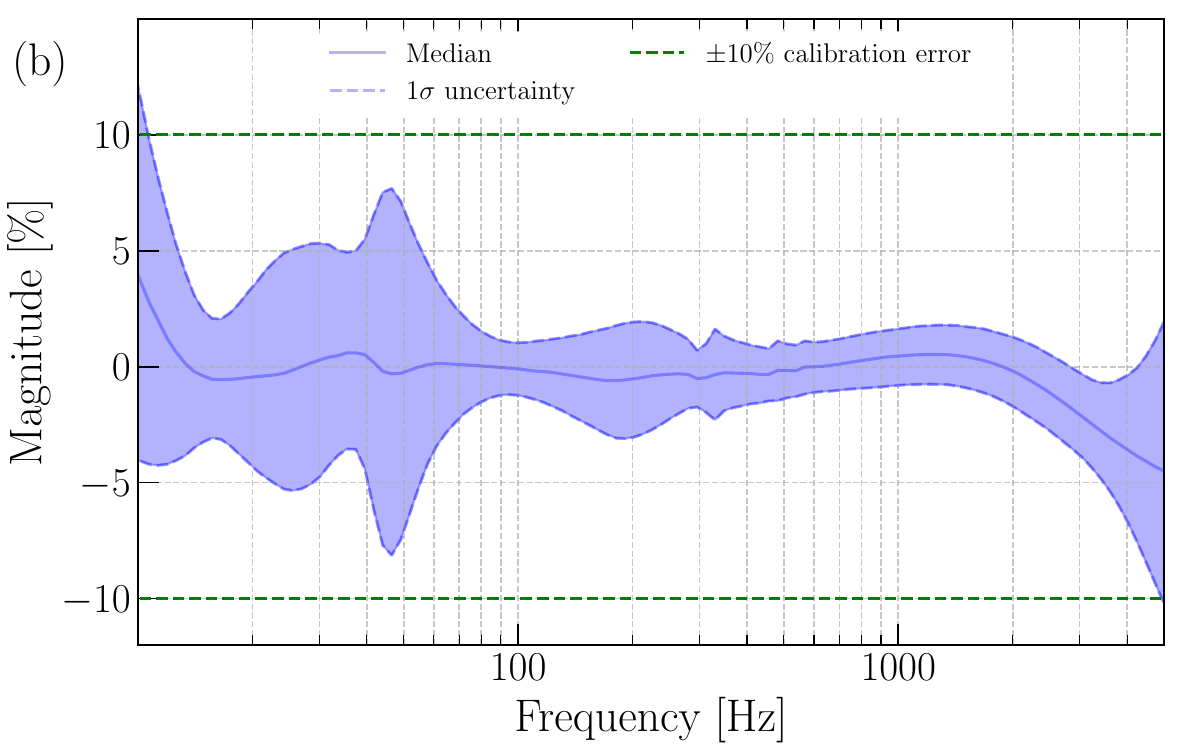}
    \includegraphics[width=0.96\linewidth]{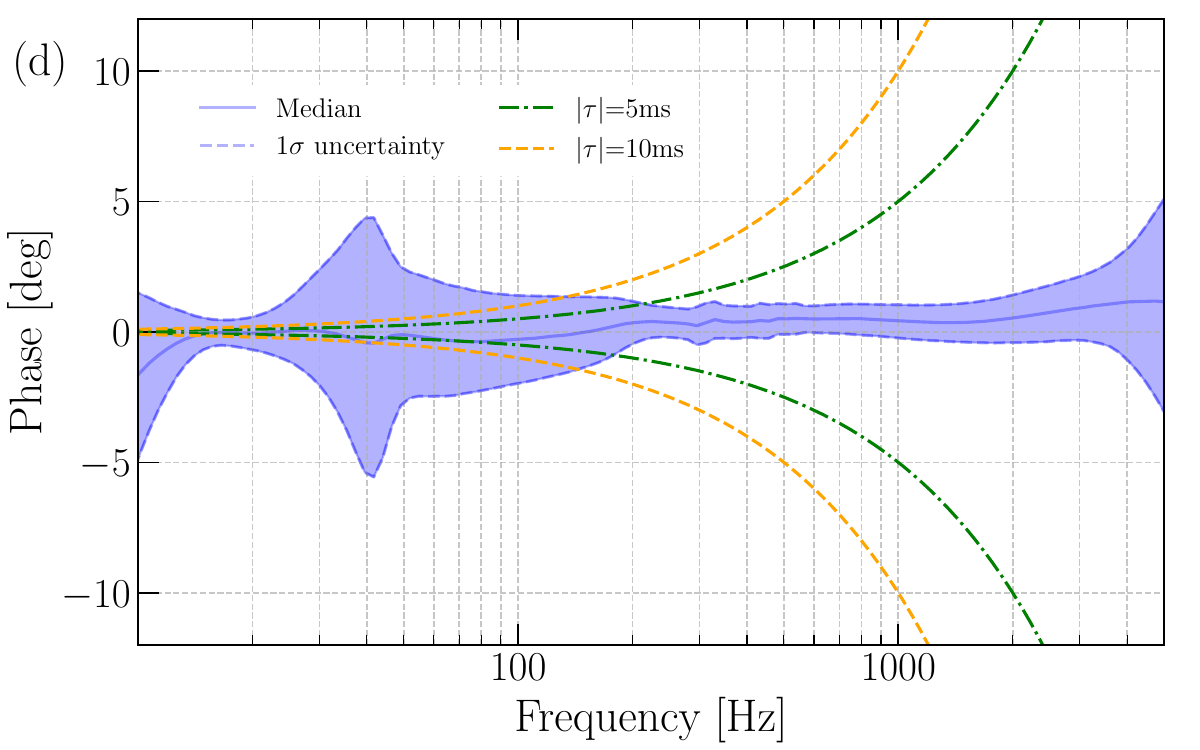}
  \end{minipage}
  \caption{\label{fig:cal}
   The physical frequency-dependent calibration errors for magnitude, panels (a) and (b), and phase, panels (c) and (d), for H1 and L1, respectively. These examples correspond to GPS times of the worst calibration errors during O3. The dashed lines in panels (a) and (b) show the amplitude calibration errors used in the previous all-sky search~\cite{c2}. The dashed lines in panels (c) and (d) show the induced phase calibration errors when using a time jittering of 5\,ms and 10\,ms as indicated by the green and orange curves, respectively. When compared to the realistic calibration curves, these two methods yield estimates for the calibration errors that are non-representative of the magnitude or frequency evolution of possible physical calibration errors. The realistic calibration errors are found to be negligible with respect to the previously used ones.
     }
%   \label{fig:eff}
\end{figure*}

The previous optically targeted searches~\cite{SNSearchS5A5S6,SNSearchO1O2} applied a conservative 9.1\% impact on detection efficiency. This value was calculated based on the all-sky search~\cite{c2} during the Fifth Science run of LIGO and Virgo (2009). Over the years, the calibration errors significantly improved, and the conservative value is revisited in this search.

The previous methods calculating the impact of calibration errors in the all-sky search~\cite{c2} used simplified models of the instrument's response. In particular, they assumed that the distortions at each interferometer would result either in delays or fluctuations of the overall amplitude (\textit{time jittering}) for narrowband GWs. In reality, for broadband GW signals like the ones produced by CCSNe, the explicit dependence of the calibration errors in amplitude and phase with respect to the frequency can result in distortion of the signal's time series independently at each interferometer. This effect can result in reduction of the cWB search parameters $\eta_{\rm c}$, $c_{\rm c}$ and detection efficiencies (see Sec.~\ref{sec:cwb}), and consequently, the search sensitivity. This Appendix describes how the systematic error and uncertainty estimates impact the search sensitivity of the analysis conducted in this paper.

Figure~\ref{fig:cal} compares an example of physically motivated worst calibration errors during O3 with the calibration errors used in the previous all-sky search~\cite{c2}. Panels (a) and (b) of Figure~\ref{fig:cal} show the frequency-dependent amplitude calibration errors for LIGO Hanford (H1) and Livingston (L1), respectively, contrasted with the dashed 10\% amplitude band used in~\cite{c2}. Panels (c) and (d) of Figure~\ref{fig:cal} show the frequency-dependent phase calibration errors for H1 and L1, respectively, contrasted with the dashed lines indicating the induced phase errors using the time jittering method used in~\cite{c2}. For both detectors, the calibration errors derived from the time jittering method are inconsistent with the frequency dependence of realistic phase and amplitude calibration errors.

Given that the search in this paper is conducted with a network of L1 and H1 detectors, the impact of calibration errors described in this Appendix is considered only for that network configuration. We do not study the impact on the networks involving the Virgo detector; however, the Virgo calibration errors can be seen in Figures~19 and~20 in Ref.~\cite{VIRGO:2021kfv}.

We modify the cWB pipeline to accept frequency-dependent amplitude and phase errors at the desired frequencies. This allows for the perturbation of the injections according to the realistic calibration errors and an estimate of the impact on the cWB detection parameters $\eta_{\rm c}$ and $c_{\rm c}$, and the detection efficiencies. The impact of calibration errors is better isolated with large signal-to-noise ratio events where the interferometric noise becomes negligible. Therefore, the impact on $\eta_{\rm c}$, $c_{\rm c}$ and detection efficiencies were quantified for waveforms injected at a distance of 1\,kpc. At closer distances, the amplitude of the injections interferes with the internal tuning of the cWB parameters. We find that the impact on $\eta_{\rm c}$, $c_{\rm c}$, and detection efficiencies at the times of the different OSWs is negligible for the network of H1 and L1 detectors.

%%%%%%%%%%%%%%%%%%%%%%%%%%%%%%%%%%%%
\bibliography{snsearch}
%%%%%%%%%%%%%%%%%%%%%%%%%%%%%%%%%%%%

\end{document}

%% file: macros.tex
\renewcommand{\today}{\number\day\space\ifcase\month\or
  January\or February\or March\or April\or May\or June\or
  July\or August\or September\or October\or November\or December\fi
  \space\number\year}

\def\be{\begin{equation}}
\def\ee{\end{equation}}
\def\bi{\begin{itemize}} 
\def\ei{\end{itemize}}
\def\ben{\begin{enumerate}}
\def\een{\end{enumerate}}

\def\hrss{h_\mathrm{rss}}

\def\aj{Astron. J.}
\def\apj{Astrophys. J.}
\def\apjl{Astrophys. J. Lett.}
\def\apjs{Astrophys. J. Supp. Ser. }

\def\aap{Astron. Astrophys. }
\def\araa{Ann.\ Rev. Astron. Astroph. }

\def\mnras{Mon. Not. Roy. Astron. Soc. }

\def\prl{Phys. Rev. Lett.}
\def\prd{Phys. Rev. D.}